\documentclass[11pt,a4paper]{article}
\pdfoutput=1
\usepackage[dvipsnames]{xcolor}
\usepackage{graphicx,caption,array}
\usepackage{ifpdf}
\usepackage{xspace}
\usepackage{jheppub}
\usepackage{dsfont,amsfonts,amsmath,euscript,amssymb,bbold}
\usepackage{xurl}
\usepackage{axodraw2}
\usepackage{multirow}
\usepackage{pstricks}
\usepackage{relsize,exscale,scalefnt,anyfontsize,cases}
\usepackage[utf8]{inputenc}
\usepackage{relsize} 
\usepackage{soul}
\usepackage{genyoungtabtikz}
\usepackage{tikz}
\usetikzlibrary{calc}

\usetikzlibrary{patterns}
\usetikzlibrary{decorations.pathmorphing,decorations.markings,trees,calc,shapes.geometric}
\tikzset{
  qua/.style={postaction={decorate},decoration={markings,mark=at position .55 with {\large\arrow{>}}}},
  anti/.style={postaction={decorate},decoration={markings,mark=at position .55 with {\large\arrow{<}}}},
  glu/.style={draw = black, decorate,decoration={coil,amplitude=1.4pt, segment length=2.6pt}} ,
  glu2/.style={draw = blue} ,
  vertical align/.style={baseline=-.5*(height("$+$")-depth("$+$"))},
  every picture/.style={vertical align}
}
 
\newcommand*{\boxcolor}{orange}
\makeatletter
\renewcommand{\boxed}[1]{\textcolor{\boxcolor}{%
\tikz[baseline={([yshift=-1ex]current bounding box.center)}] \node [rectangle, minimum width=1ex,rounded corners,draw] {\normalcolor\m@th$\displaystyle#1$};}}
 \makeatother

\def\ed{\end{document}}
\def\bc{\begin{center}}
\def\ec{\end{center}}
\def\bi{\begin{itemize}}
\def\ei{\end{itemize}}
\def\ben{\begin{enumerate}}
\def\een{\end{enumerate}}
\def\be{\begin{equation}}
\def\ee{\end{equation}}
\def\bea{\begin{eqnarray}}
\def\eea{\end{eqnarray}}
\newcommand{\nn}{\nonumber} 
\newcommand{\ie}{{i.e.}}  
\newcommand{\eg}{{e.g.}}
\def\bex{\begin{exercise}}
\def\eex{\end{exercise}}
\def\ba{\begin{array}}
\def\ea{\end{array}}
\def\bs{\begin{scope}}
\def\es{\end{scope}}

\def\sun{{\rm SU}(N)}

\newcommand{\unit}{{\mathds{1}}}
\newcommand{\Nc}{N}
\newcommand{\CF}{C_{_{\rm F}}}

\newcommand{\eq}[1]{(\ref{#1})}

\newcommand{\al}{\alpha}

\def\bm#1{\mbox{\boldmath$#1$}}
\newcommand{\tr}{\mathrm{Tr}\,}

\newcommand{\mathsize}[2]{\mbox{\fontsize{#1}{2}\selectfont $#2$}}

\newcommand{\Proj}[1]{{P}_{\mathsize{8}{\! #1}}}

\newcommand{\Irrep}[2]{{\bm{#1}}_{#2}}

\def \drawqua (#1) {
\draw[thick,postaction={decorate},decoration={markings,mark=at position #1 with {\large\arrow{>}}}] 
}
\def \drawanti (#1) {
\draw[thick,postaction={decorate},decoration={markings,mark=at position #1 with {\large\arrow{<}}}] 
}

\usepackage{tikz}
\usetikzlibrary{patterns}
\usetikzlibrary{decorations.pathmorphing,decorations.markings,trees,calc,shapes.geometric}
\tikzset{
  qua/.style={postaction={decorate},decoration={markings,mark=at position .55 with {\large\arrow{>}}}},
  anti/.style={postaction={decorate},decoration={markings,mark=at position .55 with {\large\arrow{<}}}},
  glu/.style={decorate,decoration={coil,amplitude=1.4pt, segment length=2.6pt}} ,
  glu2/.style={draw = blue} ,
  vertical align/.style={baseline=-.5*(height("$+$")-depth("$+$"))},
  every picture/.style={vertical align}
}

\newcommand{\tikeq}[1]{\vcenter{\hbox{\begin{tikzpicture}#1\end{tikzpicture}}}}
\newcommand{\tikeqbis}[1]{\hbox{\begin{tikzpicture}#1\end{tikzpicture}}}

\newcommand{\Idq}[1]{\tikeqbis{\begin{scope}[scale=#1]
\drawqua(0.55) (0,0) -- (0.8,0);
\end{scope}
}}

\newcommand{\Idg}[1]{\tikeqbis{\begin{scope}[scale=#1]
    \draw[thick,glu] (0,0) -- ++(0.93,0);
\end{scope}
}}

\newcommand{\Idqq}[1]{\tikeq{\begin{scope}[scale=#1]
\drawqua(0.55) (0,0.2) -- (0.8,0.2); 
\drawqua(0.55) (0,-0.2) -- (0.8,-0.2); 
\end{scope}
}}  

\newcommand{\projblob}[2]{\tikeq{\begin{scope}[scale=#2]
\drawqua(0.6) (0,0.2) -- (0.5,0.2);  
\drawqua(0.6) (0,-0.2) -- (0.5,-0.2); 
\drawqua(0.6) (1,0.2) -- (1.5,0.2); 
\drawqua(0.6) (1,-0.2) -- (1.5,-0.2); 
\draw[fill=white] (0.75,0) ellipse (3mm and 4mm);
\node at (0.75,0) {$#1$};
\end{scope}}}

\newcommand{\Xqq}[2]{\tikeqbis{\begin{scope}[scale=#1,yshift=#2]
\drawqua(0.3) (0,0.2) -- (0.8,-0.2); 
\drawqua(0.3) (0,-0.2) -- (0.8,0.2);
\end{scope}
}}

\newcommand{\PS}[1]{
\tikeq{
\begin{scope}[scale=#1]
\drawqua(0.5) (0,0.2) -- (1,0.2); \drawqua(0.5) (0.2,0.2) -- (0.3,0.2); \drawqua(0.5) (0.8,0.2) -- (0.85,0.2); 
\drawqua(0.5) (0,-0.2) -- (1,-0.2); \drawqua(0.5) (0.2,-0.2) -- (0.3,-0.2); \drawqua(0.5) (0.8,-0.2) -- (0.85,-0.2); 
\draw[fill=white,opacity=1] (0.4,-0.3) rectangle (0.6,0.3); 
\end{scope}
}}

\newcommand{\PA}[1]{
\tikeq{
\begin{scope}[scale=#1]
\drawqua(0.5) (0,0.2) -- (1,0.2); \drawqua(0.5) (0.2,0.2) -- (0.3,0.2); \drawqua(0.5) (0.8,0.2) -- (0.85,0.2); 
\drawqua(0.5) (0,-0.2) -- (1,-0.2); \drawqua(0.5) (0.2,-0.2) -- (0.3,-0.2); \drawqua(0.5) (0.8,-0.2) -- (0.85,-0.2); 
\draw[fill=black,opacity=1] (0.4,-0.3) rectangle (0.6,0.3); 
\end{scope}
}}

\newcommand{\ampblob}[2]{\tikeq{\begin{scope}[scale=#2]
\drawqua(0.6) (0,0.2) -- (0.5,0.2);  
\drawqua(0.6) (0,-0.2) -- (0.5,-0.2); 
\drawqua(0.6) (1.3,0.2) -- (1.8,0.2); 
\drawqua(0.6) (1.3,-0.2) -- (1.8,-0.2); 
\draw[fill=white] (0.9,0) ellipse (4.5mm and 5.5mm);
\node at (0.9,0) {$#1$};
\end{scope}}}

\newcommand{\Mblob}[2]{\tikeq{\begin{scope}[scale=#2]
\draw[fill=white] (0.9,0) ellipse (4.5mm and 5.5mm);
\node at (0.9,0) {$#1$};
\end{scope}}}

\newcommand{\Casimirop}[2]{\tikeqbis{\begin{scope}[scale=#2]
\drawqua(0.25) (0,0.2) -- (1,0.2); 
\drawqua(0.25) (0,-0.2) -- (1,-0.2); 
\drawqua(0.85) (1.5,0.2) -- (2.5,0.2); 
\drawqua(0.85) (1.5,-0.2) -- (2.5,-0.2); 
\draw[fill=white] (1.25,0) ellipse (4.5mm and 5.5mm);
\node at (1.25,0) {$#1$};
\draw[glu] (0.5,0.2) .. controls +(0,0.5) and +(-0.5,0) .. (1.25,0.9) 
.. controls +(0.5,0) and +(0,0.5) .. (2,0.2); 
\draw[fill=white!70!black] (0.5,0) ellipse (.1 and .4);
\draw[fill=white!70!black] (2,0) ellipse (.1 and .4);
\end{scope}}}

\newcommand{\CasPPinside}[1]{\tikeqbis{\begin{scope}[scale=#1]
\draw[thick] (-0.2,0.2) -- (2.7,0.2); 
\drawqua(0.6) (-0.2,0.2) -- (0.2,0.2); 
\drawqua(0.6) (2.3,0.2) -- (2.7,0.2); 
\draw[thick] (-0.2,-0.2) -- (2.7,-0.2); 
\drawqua(0.6) (-0.2,-0.2) -- (0.2,-0.2); 
\drawqua(0.6) (2.3,-0.2) -- (2.7,-0.2); 
\draw[fill=white] (0.8,0) ellipse (2.5mm and 3.5mm);
\node at (0.8,0) {$\al$};
\draw[fill=white] (1.7,0) ellipse (2.5mm and 3.5mm);
\node at (1.7,0) {$\al$};
\draw[glu] (0.3,0.2) .. controls +(0,0.6) and +(-0.5,0) .. (1.25,0.9) 
.. controls +(0.5,0) and +(0,0.6) .. (2.2,0.2); 
\draw[fill=white!70!black] (0.3,0) ellipse (.1 and .4);
\draw[fill=white!70!black] (2.2,0) ellipse (.1 and .4);
\end{scope}}}

\newcommand{\CasPPoutside}[1]{\tikeqbis{\begin{scope}[scale=#1]
\draw[thick] (-0.2,0.2) -- (2.7,0.2); 
\drawqua(0.6) (-0.2,0.2) -- (0.2,0.2); 
\drawqua(0.65) (2.3,0.2) -- (2.7,0.2); 
\draw[thick] (-0.2,-0.2) -- (2.7,-0.2); 
\drawqua(0.6) (-0.2,-0.2) -- (0.2,-0.2); 
\drawqua(0.65) (2.3,-0.2) -- (2.7,-0.2); 
\draw[fill=white] (0.4,0) ellipse (2.5mm and 3.5mm);
\node at (0.4,0) {$\al$};
\draw[fill=white] (2.1,0) ellipse (2.5mm and 3.5mm);
\node at (2.1,0) {$\al$};
\draw[glu] (1,0.2) .. controls +(0,0.5) and +(-0.15,0) .. (1.25,0.9) 
.. controls +(0.15,0) and +(0,0.5) .. (1.5,0.2); 
\draw[fill=white!70!black] (1,0) ellipse (.1 and .4);
\draw[fill=white!70!black] (1.5,0) ellipse (.1 and .4);
\end{scope}}}

\newcommand{\CasCal}[1]{\tikeqbis{\begin{scope}[scale=#1]
\draw[thick] (0,0.2) -- (1.6,0.2); 
\drawqua(0.65) (0,0.2) -- (0.3,0.2); 
\drawqua(0.65) (1.3,0.2) -- (1.6,0.2); 
\draw[thick] (0,-0.2) -- (1.6,-0.2); 
\drawqua(0.65) (0,-0.2) -- (0.3,-0.2); 
\drawqua(0.65) (1.3,-0.2) -- (1.6,-0.2); 
\draw[glu] (0.4,0.2) .. controls +(0,0.5) and +(-0.1,0) .. (0.8,0.7) 
.. controls +(0.1,0) and +(0,0.5) .. (1.2,0.2); 
\draw[fill=white!70!black] (0.4,0) ellipse (.1 and .4);
\draw[fill=white!70!black] (1.2,0) ellipse (.1 and .4);
\end{scope}}}

\newcommand{\Vqq}[2]{\tikeqbis{\bs[scale=#1,yshift=#2]
\draw[thick,qua] (0,-0.2) -- (0.4,-0.2);
\draw[thick,qua] (0.4,-0.2) -- (0.8,-0.2);
\draw[thick,qua] (0,0.2) -- (0.4,0.2);
\draw[thick,qua] (0.4,0.2) -- (0.8,0.2);
\draw[thick,glu] (0.4,-0.2) -- (0.4,0.2);
\es
}}

\newcommand{\Fierzmapa}[1]{\tikeq{\begin{scope}[scale=#1] 
\drawqua(0.6) (0,-0.2) -- ++(0.8,0);
\drawanti(0.5) (0,.2) -- ++(0.8,0); 
\end{scope} 
}}	  

\newcommand{\Fierzmapb}[1]{\tikeq{\begin{scope}[scale=#1] 
\drawqua(0.55) (0,-.25) -- ++(0.2,0) -- ++(0,.5) -- ++(-0.2,0);  
\drawanti(0.5) (0.6,-.25) -- ++(-0.2,0) -- ++(0,.5) -- ++(0.2,0); 
\end{scope} 
}} 

\newcommand{\Fierzmapc}[1]{\tikeq{\begin{scope}[scale=#1] 
\drawqua(0.6) (0,-.3) -- ++(.3,.3);
\drawanti(0.5) (0,.3) -- ++(.3,-.3);
\draw[thick,glu] (.3,0) -- ++(.6,0);
\drawanti(0.6) (.9,0) -- ++(.3,.3);
\drawqua(0.7) (.9,0) -- ++(.3,-.3); 
\end{scope} 
}}

\newcommand{\ColorConservationA}[1]{\tikeqbis{\bs[scale=#1]
 \draw[thick,->] (3,2) -- ++(1,0); \draw[thick] (4,2) -- ++(1,0); 
 \draw[thick] (3,2) to[out=180,in=45] (0,1);
 \draw[thick,-<] (3,1) -- ++(1,0); \draw[thick] (3,1) -- ++(2,0);
 \draw[thick] (3,1) to[out=180,in=30] (0,.5);
 \draw[glu,thick] (0,0) to (3,0) -- ++(2,0);
 \draw[glu,thick] (0,-1) to[out=-45,in=180] (3,-2) -- ++(2,0);
 \draw[fill=white,opacity=1] (0,0) circle (2); 
 \draw[fill=black] (4.5,-.5) circle (0.05);
 \draw[fill=black] (4.5,-1) circle (0.05);
 \draw[fill=black] (4.5,-1.5) circle (0.05);
 \draw[glu,thick] (3,-4.1) -- (3,2);
\es
}}

\newcommand{\ColorConservationB}[1]{\tikeqbis{\bs[scale=#1]
 \draw[thick,->] (3,2) -- ++(1,0); \draw[thick] (4,2) -- ++(1,0); 
 \draw[thick] (3,2) to[out=180,in=45] (0,1);
 \draw[thick,-<] (3,1) -- ++(1,0); \draw[thick] (3,1) -- ++(2,0);
 \draw[thick] (3,1) to[out=180,in=30] (0,.5);
 \draw[glu,thick] (0,0) to (3,0) -- ++(2,0);
 \draw[glu,thick] (0,-1) to[out=-45,in=180] (3,-2) -- ++(2,0);
 \draw[fill=white,opacity=1] (0,0) circle (2); 
 \draw[fill=black] (4.5,-.5) circle (0.05);
 \draw[fill=black] (4.5,-1) circle (0.05);
 \draw[fill=black] (4.5,-1.5) circle (0.05);
\draw[glu,thick] (3,-4.1) -- (3,1);
\es
}}

\newcommand{\ColorConservationC}[1]{\tikeqbis{\bs[scale=#1]
 \draw[thick,->] (3,2) -- ++(1,0); \draw[thick] (4,2) -- ++(1,0); 
 \draw[thick] (3,2) to[out=180,in=45] (0,1);
 \draw[thick,-<] (3,1) -- ++(1,0); \draw[thick]  (3,1) -- ++(2,0);
 \draw[thick] (3,1) to[out=180,in=30] (0,.5);
 \draw[glu,thick] (0,0) to (3,0) -- ++(2,0);
 \draw[glu,thick] (0,-1) to[out=-45,in=180] (3,-2) -- ++(2,0);
 \draw[fill=white,opacity=1] (0,0) circle (2); 
 \draw[fill=black] (4.5,-.5) circle (0.05);
 \draw[fill=black] (4.5,-1) circle (0.05);
 \draw[fill=black] (4.5,-1.5) circle (0.05);
 \draw[glu,thick] (3.1,-4.1) -- (3.1,0.1)node{$\scriptstyle\black\bullet$};
\es
}}

\newcommand{\ColorConservationD}[1]{\tikeqbis{\bs[scale=#1]
 \draw[thick,->] (3,2) -- ++(1,0); \draw[thick] (4,2) -- ++(1,0); 
 \draw[thick] (3,2) to[out=180,in=45] (0,1);
 \draw[thick,-<] (3,1) -- ++(1,0); \draw[thick] (3,1) -- ++(2,0);
 \draw[thick] (3,1) to[out=180,in=30] (0,.5);
 \draw[glu,thick] (0,0) to (3,0) -- ++(2,0);
 \draw[glu,thick] (0,-1) to[out=-45,in=180] (3,-2) -- ++(2,0);
 \draw[fill=white,opacity=1] (0,0) circle (2); 
 \draw[fill=black] (4.5,-.5) circle (0.05);
 \draw[fill=black] (4.5,-1) circle (0.05);
 \draw[fill=black] (4.5,-1.5) circle (0.05);
 \draw[glu,thick] (3.0,-4.1) -- (3.0,-2)node{$\scriptstyle\black\bullet$};
\es
}}

\newcommand{\colconsLeft}[1]{\tikeqbis{\begin{scope}[scale=#1] 
\draw[qua,thick] (-2.2,0.5) -- (-0.4,0.5); 
\draw[anti,thick] (-2.2,0) -- (-0.4,0); 
\draw[glu,thick] (-2.2,-0.5) -- (-0.4,-0.5) ; 
\draw[glu,thick] (0.6,0.5) -- (2.05,0.5); 
\draw[glu,thick] (0.6,0.17) -- (2.05,0.17);
\draw[glu,thick] (0.6,-0.17) -- (2.05,-0.17);    
\draw[glu,thick] (0.6,-0.5) -- (2.05,-0.5); 
\draw[fill=white,opacity=1] (0,0) circle (0.8); 
\draw (0,0) node{$A$} ;
\draw[glu,thick] (-1.7,-1.3) -- (-1.7,-0.75); 
\draw[fill=white!70!black] (-1.7,0.7) .. controls +(-0.2,0) and +(-0.2,0) .. (-1.7,-0.75); 
\draw[fill=white!70!black] (-1.7,0.7) .. controls +(0.2,0) and +(0.2,0) .. (-1.7,-0.75);
\end{scope}
}}

\newcommand{\colconsRight}[1]{\tikeqbis{\begin{scope}[scale=#1] 
\draw[qua,thick] (-2,0.5) -- (-0.6,0.5); 
\draw[anti,thick] (-2,0) -- (-0.6,0); 
\draw[glu,thick] (-2,-0.5) -- (-0.6,-0.5) ; 
\draw[glu,thick] (0.6,0.5) -- (2.05,0.5); 
\draw[glu,thick] (0.6,0.17) -- (2.05,0.17);
\draw[glu,thick] (0.6,-0.17) -- (2.05,-0.17);
\draw[glu,thick] (0.6,-0.5) -- (2.05,-0.5); 
\draw[fill=white,opacity=1] (0,0) circle (0.8); 
\draw (0,0) node{$A$} ;
\draw[glu,thick] (1.55,-1.3) -- (1.55,-0.75); 
\draw[fill=white!70!black]  (1.55,0.7) .. controls +(-0.2,0) and +(-0.2,0) .. (1.55,-0.75); 
\draw[fill=white!70!black]  (1.55,0.7) .. controls +(0.2,0) and +(0.2,0) .. (1.55,-0.75);
\end{scope}
}}
 
\def \gqqVert (#1) {
\tikeq{\begin{scope}[scale=#1]
   \draw (0,0) node[left] {$i$};
  \draw[thick,qua](0,0) -- (.4,0);
  \draw[thick,qua](.4,0) -- (.8,0);
  \draw (.8,0) node[right] {$j$};
  \draw[thick,glu](.4,-0.4) -- (.4,0);
  \draw (.4,-0.4) node[below] {$a$};
    \end{scope}}}
  
 \def \gqbarqbarVert (#1) {
\tikeq{\begin{scope}[scale=#1]  
   \draw (0,0) node[left] {$i$};
  \draw[thick,anti](0,0) -- (.4,0);
  \draw[thick,anti](.4,0) -- (.8,0);
  \draw (.8,0) node[right] {$j$};
  \draw[thick,glu](.4,-0.4) -- (.4,0);
  \draw (.4,-0.4) node[below] {$a$};
   \end{scope}}}
   
\def \gggVert (#1) {
\tikeq{\begin{scope}[scale=#1]   
   \draw (0,0) node[left] {$c$};
  \draw[thick,glu](0,0) -- (.4,0)node{$\scriptstyle\bullet$};
  \draw[thick,glu](.4,0) -- (.8,0);
  \draw (.8,0) node[right] {$b$};
  \draw[thick,glu](.4,-0.4) -- (.4,0);
  \draw (.4,-0.4) node[below] {$a$};
 \end{scope}}}
 
 \newcommand{\trtt}[1]{\tikeq{\bs[scale=#1]
 \drawanti(0.25) (0,0) circle (2);
 \draw[thick,glu] (2,0) -- (4.5,0); \draw[thick,glu] (-4.5,0) -- (-2,0);
\es
}}

\newcommand{\projblobmulti}[2]{\tikeq{\begin{scope}[scale=#2]
\drawqua(0.6) (0,0.4) node[left] {$\mathsize{6}{1}$} -- (0.5,0.4);  
\drawqua(0.6) (0,0.2) node[left] {$\mathsize{6}{2}$} -- (0.5,0.2);  
\drawqua(0.6) (0,-0.4)  -- (0.5,-0.4); 
\node at (-0.25,-0.4) {$\mathsize{6}{b\!\!-\!\!1}$}; 
\node at (0.1,-0.) {$\mathsize{11}{\vdots}$}; 
\node at (1.4,-0.) {$\mathsize{11}{\vdots}$}; 
\drawqua(0.6) (1,0.4) -- (1.5,0.4); 
\drawqua(0.6) (1,0.2) -- (1.5,0.2); 
\drawqua(0.6) (1,-0.4) -- (1.5,-0.4); 
\draw[fill=white] (0.75,0) ellipse (3.5mm and 6mm);
\node at (0.75,0) {$#1$};
\end{scope}}}

\newcommand{\Idmult}[2]{\tikeqbis{\begin{scope}[scale=#2,yshift=2mm]
\drawqua(0.6) (0,0.4) node[left] {$\mathsize{6}{1}$} -- (0.5,0.4);  
\drawqua(0.6) (0,0.2) node[left] {$\mathsize{6}{2}$} -- (0.5,0.2);  
\drawqua(0.6) (0,-0.4)  -- (0.5,-0.4); 
\node at (-0.25,-0.4) {$\mathsize{6}{b\!\!-\!\!1}$}; 
\node at (0.1,-0.) {$\mathsize{11}{\vdots}$}; 
\node at (1.4,-0.) {$\mathsize{11}{\vdots}$}; 
\drawqua(0.6) (1,0.4) -- (1.5,0.4); 
\drawqua(0.6) (1,0.2) -- (1.5,0.2); 
\drawqua(0.6) (1,-0.4) -- (1.5,-0.4); 
\drawqua(0.6) (0,-0.8) -- (1.5,-0.8); 
\draw[fill=white] (0.75,0) ellipse (3.5mm and 6mm);
\node at (0.75,0) {$#1$};
\end{scope}}}

\newcommand{\projmultalbeta}[3]{\tikeqbis{
\begin{scope}[scale=#3,yshift=2mm]
\drawqua(0.6) (0,0.4) -- (0.5,0.4); 
\drawqua(0.6) (0,0.2)  -- (0.5,0.2); 
\drawqua(0.6) (0,-0.4)  -- (0.5,-0.4); 
\node at (0.1,-0.) {$\mathsize{11}{\vdots}$}; 
\node at (1.4,-0.) {$\mathsize{11}{\vdots}$}; 
\drawqua(0.6) (1,0.4) -- (1.5,0.4); 
\drawqua(0.6) (1,0.2) -- (1.5,0.2); 
\drawqua(0.6) (1,-0.4) -- (1.5,-0.4); 
\drawqua(0.6) (0,-0.8) -- (1.5,-0.8); 
\draw[fill=white] (0.75,0) ellipse (3.5mm and 6mm);
\node at (0.75,0) {$#1$};
\end{scope}
\begin{scope}[scale=#3,yshift=2mm,xshift=20.2mm]
\drawqua(0.6) (0,0.4) -- (0.5,0.4); 
\drawqua(0.6) (0,0.2)  -- (0.5,0.2); 
\drawqua(0.6) (0,-0.4)  -- (0.5,-0.4); 
\node at (0.1,-0.) {$\mathsize{11}{\vdots}$}; 
\node at (1.4,-0.) {$\mathsize{11}{\vdots}$}; 
\drawqua(0.6) (1,0.4) -- (1.5,0.4); 
\drawqua(0.6) (1,0.2) -- (1.5,0.2); 
\drawqua(0.6) (1,-0.4) -- (1.5,-0.4); 
\drawqua(0.6) (0,-0.8) -- (1.5,-0.8); 
\draw[fill=white] (0.75,0) ellipse (3.5mm and 6mm);
\node at (0.75,0) {$#1$};
\end{scope}
\begin{scope}[scale=#3,yshift=2mm,xshift=-2.4mm]
\draw[fill=white] (2,-0.2) ellipse (4mm and 8mm);
\node at (2,-0.2) {$#2$};
\end{scope}
}}

\newcommand{\projmultalbetacut}[3]{\tikeqbis{
\begin{scope}[scale=#3,yshift=2mm]
\drawqua(0.6) (0,0.4) -- (0.5,0.4); 
\drawqua(0.6) (0,0.2)  -- (0.5,0.2); 
\drawqua(0.6) (0,-0.4)  -- (0.5,-0.4); 
\node at (0.1,-0.) {$\mathsize{11}{\vdots}$}; 
\node at (1.25,-0.) {$\mathsize{11}{\vdots}$}; 
\drawqua(0.6) (1,0.4) -- (1.5,0.4); 
\drawqua(0.6) (1,0.2) -- (1.5,0.2); 
\drawqua(0.6) (1,-0.4) -- (1.5,-0.4); 
\drawqua(0.2) (0,-0.8) -- (1.5,-0.8); 
\draw[fill=white] (0.75,0) ellipse (3.5mm and 6mm);
\node at (0.75,0) {$#1$};
\drawqua(0.6) (2.5,0.4) -- (3,0.4); 
\drawqua(0.6) (2.5,0.2)  -- (3,0.2); 
\drawqua(0.6) (2.5,-0.4)  -- (3,-0.4); 
\node at (2.9,-0.) {$\mathsize{11}{\vdots}$}; 
\drawqua(0.6) (2.5,-0.8) -- (3,-0.8); 
\draw[thick] (1.5,0.4) -- (2.5,0.4); 
\draw[thick] (1.5,0.2) -- (2.5,0.2); 
\draw[thick] (1.5,-0.4) -- (2.5,-0.4); 
\draw[thick] (1.5,-0.8) -- (2.5,-0.8); 
\draw[fill=white] (2.03,-0.2) ellipse (6.5mm and 8mm);
\node at (2.03,-0.2) {$\mathsize{11}{#2}$};
\end{scope}
}}

\newcommand{\Xmult}[1]{\tikeqbis{\begin{scope}[scale=#1,yshift=2mm]
\drawqua(0.55) (0,0.4) node[left] {$\mathsize{6}{1}$} -- (1.5,0.4);  
\drawqua(0.6) (0,-0.05) node[left] {$\mathsize{6}{k}$} -- (0.5,-0.05);  
\drawqua(0.6) (0,-0.5)  -- (0.5,-0.5); 
\drawqua(0.6) (1,-0.5)  -- (1.5,-0.5); 
\draw[thick] (0.5,-0.5) -- (1,-0.5) ; 
\node at (-0.23,-0.5) {$\mathsize{6}{b\!\!-\!\!1}$}; 
\node at (0.1,-0.18) {$\mathsize{11}{\vdots}$}; 
\node at (1.4,-0.18) {$\mathsize{11}{\vdots}$}; 
\node at (0.1,0.26) {$\mathsize{11}{\vdots}$}; 
\node at (1.4,0.26) {$\mathsize{11}{\vdots}$}; 
\drawqua(0.6) (1,-0.05) -- (1.5,-0.05); 
\drawqua(0.6) (0,-0.8) -- (0.5,-0.8); 
\drawqua(0.6) (1,-0.8) -- (1.5,-0.8); 
\draw[thick] (0.5,-0.05)  .. controls +(0.25,0) and +(-0.2,0) ..  (1,-0.8);
\draw[thick] (0.5,-0.8) .. controls +(0.2,0) and +(-0.25,0) ..  (1,-0.05);
\end{scope}}}

\newcommand{\projthree}[4]{\tikeqbis{
\begin{scope}[xscale=#3,yscale=#4,yshift=0mm]
\drawqua(0.1) (0,0.4) -- (1.65,0.4); 
\drawqua(0.1) (0,0)  -- (1.65,0); 
\drawqua(0.1) (0,-0.4) -- (1.65,-0.4); 
\draw[fill=white] (0.45,0.2) ellipse (2mm and 4mm); 
\node at (0.45,0.2) {$\mathsize{8}{#1}$};
\draw[fill=white] (1.1,0) ellipse (3mm and 6mm);
\node at (1.1,0) {$\mathsize{8}{#2}$};
\end{scope}
}}

\newcommand{\projfour}[5]{\tikeq{
\begin{scope}[xscale=#4,yscale=#5,yshift=0mm]
\drawqua(0.06) (0,0.4) -- (2.5,0.4); 
\drawqua(0.06) (0,0)  -- (2.5,0); 
\drawqua(0.06) (0,-0.4) -- (2.5,-0.4); 
\drawqua(0.06) (0,-0.8) -- (2.5,-0.8); 
\draw[fill=white] (0.45,0.2) ellipse (2mm and 4mm); 
\node at (0.45,0.2) {$\mathsize{8}{#1}$};
\draw[fill=white] (1.1,0) ellipse (3mm and 6mm);
\node at (1.1,0) {$\mathsize{8}{#2}$};
\draw[fill=white] (1.9,-0.2) ellipse (3mm and 8mm);
\node at (1.9,-0.2) {$\mathsize{8}{#3}$};
\end{scope}
}}

\newcommand{\projeight}[9]{\tikeq{
\begin{scope}[xscale=#8,yscale=#9,yshift=0mm,rotate=0]
\drawqua(0.03) (0,0.4) -- (6,0.4); 
\drawqua(0.03) (0,0)  -- (6,0); 
\drawqua(0.03) (0,-0.4) -- (6,-0.4); 
\drawqua(0.03) (0,-0.8) -- (6,-0.8); 
\drawqua(0.03) (0,-1.2) -- (6,-1.2); 
\drawqua(0.03) (0,-1.6) -- (6,-1.6); 
\drawqua(0.03) (0,-2) -- (6,-2); 
\drawqua(0.03) (0,-2.4) -- (6,-2.4); 
\draw[fill=white] (0.45,0.2) ellipse (2mm and 4mm); 
\node at (0.45,0.2) {$\mathsize{8}{#1}$};
\draw[fill=white] (1,0) ellipse (2mm and 6mm);
\node at (1,0) {$\mathsize{8}{#2}$};
\draw[fill=white] (1.7,-0.2) ellipse (3.5mm and 8mm);
\node at (1.7,-0.2) {$\mathsize{8}{#3}$};
\draw[fill=white] (2.55,-0.4) ellipse (3.5mm and 10mm);
\node at (2.55,-0.4) {$\mathsize{8}{#4}$};
\draw[fill=white] (3.45,-0.6) ellipse (3.5mm and 12mm);
\node at (3.45,-0.6) {$\mathsize{8}{#5}$};
\draw[fill=white] (4.35,-0.8) ellipse (3.5mm and 14mm);
\node at (4.35,-0.8) {$\mathsize{8}{#6}$};
\draw[fill=white] (5.3,-1) ellipse (4mm and 16mm);
\node at (5.3,-1) {$\mathsize{8}{#7}$};
\end{scope}
}}

\newcommand{\pMOLDeight}[2]{\tikeq{
\begin{scope}[scale=0.88]
			\drawqua(0.5) (0.0,1.4) -- (0.6,1.4);
			\draw (0.8,1.4) -- (1.4,1.4);
			\draw (1.6,1.4) -- (2.2,1.4);
			\draw (2.4,1.4) -- (3.0,1.4);
			\draw (3.2,1.4) -- (3.8,1.4);
			\draw (4.0,1.4) -- (4.6,1.4);
			\draw (4.8,1.4) -- (5.4,1.4);
			\draw (5.6,1.4) -- (6.2,1.4);
			\draw (6.4,1.4) -- (7.0,1.4);
			\draw (7.2,1.4) -- (7.8,1.4);
			\draw (8.0,1.4) -- (8.6,1.4);
			\draw (8.8,1.4) -- (9.4,1.4);
			\draw (9.6,1.4) -- (10.2,1.4);
			\draw (10.4,1.4) -- (11.0,1.4);
			
			\drawqua(0.5) (0.0,1.0) -- (0.6,1.0);
			\draw (0.8,1.0)
			.. controls (0.98,1.0) and (1.22,0.6)
			.. (1.4,0.6);
			\draw (1.6,0.6)
			.. controls (1.78,0.6) and (2.02,1.0)
			.. (2.2,1.0);
			\draw (2.4,1.0)
			.. controls (2.58,1.0) and (2.82,0.6)
			.. (3.0,0.6);
			\draw (3.2,0.6)
			.. controls (3.38,0.6) and (3.62,1.0)
			.. (3.8,1.0);
			\draw (4.0,1.0)
			.. controls (4.18,1.0) and (4.42,0.2)
			.. (4.6,0.2);
			\draw (4.8,0.2)
			.. controls (4.98,0.2) and (5.22,1.0)
			.. (5.4,1.0);
			\draw (5.6,1.0)
			.. controls (5.78,1.0) and (6.02,0.2)
			.. (6.2,0.2);
			\draw (6.4,0.2)
			.. controls (6.58,0.2) and (6.82,1.0)
			.. (7.0,1.0);
			\draw (7.2,1.0)
			.. controls (7.38,1.0) and (7.62,0.6)
			.. (7.8,0.6);
			\draw (8.0,0.6)
			.. controls (8.18,0.6) and (8.42,1.0)
			.. (8.6,1.0);
			\draw (8.8,1.0)
			.. controls (8.98,1.0) and (9.22,0.6)
			.. (9.4,0.6);
			\draw (9.6,0.6)
			.. controls (9.78,0.6) and (10.02,1.0)
			.. (10.2,1.0);
			\draw (10.4,1.0) -- (11.0,1.0);
			
			\drawqua(0.5) (0.0,0.6) -- (0.6,0.6);
			\draw (0.8,0.6)
			.. controls (0.98,0.6) and (1.22,1.0)
			.. (1.4,1.0);
			\draw (1.6,1.0)
			.. controls (1.78,1.0) and (2.02,0.2)
			.. (2.2,0.2);
			\draw (2.4,0.2)
			.. controls (2.58,0.2) and (2.82,1.0)
			.. (3.0,1.0);
			\draw (3.2,1.0)
			.. controls (3.38,1.0) and (3.62,-0.6)
			.. (3.8,-0.6);
			\draw (4.0,-0.6)
			.. controls (4.18,-0.6) and (4.42,1.0)
			.. (4.6,1.0);
			\draw (4.8,1.0)
			.. controls (4.98,1.0) and (5.22,-0.6)
			.. (5.4,-0.6);
			\draw (5.6,-0.6)
			.. controls (5.78,-0.6) and (6.02,1.0)
			.. (6.2,1.0);
			\draw (6.4,1.0)
			.. controls (6.58,1.0) and (6.82,-0.6)
			.. (7.0,-0.6);
			\draw (7.2,-0.6)
			.. controls (7.38,-0.6) and (7.62,1.0)
			.. (7.8,1.0);
			\draw (8.0,1.0)
			.. controls (8.18,1.0) and (8.42,0.2)
			.. (8.6,0.2);
			\draw (8.8,0.2)
			.. controls (8.98,0.2) and (9.22,1.0)
			.. (9.4,1.0);
			\draw (9.6,1.0)
			.. controls (9.78,1.0) and (10.02,0.6)
			.. (10.2,0.6);
			\draw (10.4,0.6) -- (11.0,0.6);
			
			\drawqua(0.5) (0.0,0.2) -- (0.6,0.2);
			\draw (0.8,0.2) -- (1.4,0.2);
			\draw (1.6,0.2)
			.. controls (1.78,0.2) and (2.02,0.6)
			.. (2.2,0.6);
			\draw (2.4,0.6)
			.. controls (2.58,0.6) and (2.82,-0.2)
			.. (3.0,-0.2);
			\draw (3.2,-0.2)
			.. controls (3.38,-0.2) and (3.62,0.6)
			.. (3.8,0.6);
			\draw (4.0,0.6)
			.. controls (4.18,0.6) and (4.42,-0.6)
			.. (4.6,-0.6);
			\draw (4.8,-0.6)
			.. controls (4.98,-0.6) and (5.22,0.6)
			.. (5.4,0.6);
			\draw (5.6,0.6)
			.. controls (5.78,0.6) and (6.02,-0.6)
			.. (6.2,-0.6);
			\draw (6.4,-0.6)
			.. controls (6.58,-0.6) and (6.82,0.6)
			.. (7.0,0.6);
			\draw (7.2,0.6)
			.. controls (7.38,0.6) and (7.62,-0.2)
			.. (7.8,-0.2);
			\draw (8.0,-0.2)
			.. controls (8.18,-0.2) and (8.42,0.6)
			.. (8.6,0.6);
			\draw (8.8,0.6)
			.. controls (8.98,0.6) and (9.22,0.2)
			.. (9.4,0.2);
			\draw (9.6,0.2) -- (10.2,0.2);
			\draw (10.4,0.2) -- (11.0,0.2);
			
			\drawqua(0.5) (0.0,-0.2) -- (0.6,-0.2);
			\draw (0.8,-0.2) -- (1.4,-0.2);
			\draw (1.6,-0.2) -- (2.2,-0.2);
			\draw (2.4,-0.2)
			.. controls (2.58,-0.2) and (2.82,0.2)
			.. (3.0,0.2);
			\draw (3.2,0.2)
			.. controls (3.38,0.2) and (3.62,-1.0)
			.. (3.8,-1.0);
			\draw (4.0,-1.0)
			.. controls (4.18,-1.0) and (4.42,-0.2)
			.. (4.6,-0.2);
			\draw (4.8,-0.2)
			.. controls (4.98,-0.2) and (5.22,-1.0)
			.. (5.4,-1.0);
			\draw (5.6,-1.0)
			.. controls (5.78,-1.0) and (6.02,-0.2)
			.. (6.2,-0.2);
			\draw (6.4,-0.2)
			.. controls (6.58,-0.2) and (6.82,-1.0)
			.. (7.0,-1.0);
			\draw (7.2,-1.0)
			.. controls (7.38,-1.0) and (7.62,0.2)
			.. (7.8,0.2);
			\draw (8.0,0.2)
			.. controls (8.18,0.2) and (8.42,-0.2)
			.. (8.6,-0.2);
			\draw (8.8,-0.2) -- (9.4,-0.2);
			\draw (9.6,-0.2) -- (10.2,-0.2);
			\draw (10.4,-0.2) -- (11.0,-0.2);
			
			\drawqua(0.5) (0.0,-0.6) -- (0.6,-0.6);
			\draw (0.8,-0.6) -- (1.4,-0.6);
			\draw (1.6,-0.6) -- (2.2,-0.6);
			\draw (2.4,-0.6) -- (3.0,-0.6);
			\draw (3.2,-0.6)
			.. controls (3.38,-0.6) and (3.62,0.2)
			.. (3.8,0.2);
			\draw (4.0,0.2)
			.. controls (4.18,0.2) and (4.42,-1.0)
			.. (4.6,-1.0);
			\draw (4.8,-1.0)
			.. controls (4.98,-1.0) and (5.22,0.2)
			.. (5.4,0.2);
			\draw (5.6,0.2)
			.. controls (5.78,0.2) and (6.02,-1.0)
			.. (6.2,-1.0);
			\draw (6.4,-1.0)
			.. controls (6.58,-1.0) and (6.82,0.2)
			.. (7.0,0.2);
			\draw (7.2,0.2)
			.. controls (7.38,0.2) and (7.62,-0.6)
			.. (7.8,-0.6);
			\draw (8.0,-0.6) -- (8.6,-0.6);
			\draw (8.8,-0.6) -- (9.4,-0.6);
			\draw (9.6,-0.6) -- (10.2,-0.6);
			\draw (10.4,-0.6) -- (11.0,-0.6);
			
			\drawqua(0.5) (0.0,-1.0) -- (0.6,-1.0);
			\draw (0.8,-1.0) -- (1.4,-1.0);
			\draw (1.6,-1.0) -- (2.2,-1.0);
			\draw (2.4,-1.0) -- (3.0,-1.0);
			\draw (3.2,-1.0)
			.. controls (3.38,-1.0) and (3.62,-0.2)
			.. (3.8,-0.2);
			\draw (4.0,-0.2)
			.. controls (4.18,-0.2) and (4.42,-1.4)
			.. (4.6,-1.4);
			\draw (4.8,-1.4)
			.. controls (4.98,-1.4) and (5.22,-0.2)
			.. (5.4,-0.2);
			\draw (5.6,-0.2)
			.. controls (5.78,-0.2) and (6.02,-1.4)
			.. (6.2,-1.4);
			\draw (6.4,-1.4)
			.. controls (6.58,-1.4) and (6.82,-0.2)
			.. (7.0,-0.2);
			\draw (7.2,-0.2)
			.. controls (7.38,-0.2) and (7.62,-1.0)
			.. (7.8,-1.0);
			\draw (8.0,-1.0) -- (8.6,-1.0);
			\draw (8.8,-1.0) -- (9.4,-1.0);
			\draw (9.6,-1.0) -- (10.2,-1.0);
			\draw (10.4,-1.0) -- (11.0,-1.0);
			
			\drawqua(0.5) (0.0,-1.4) -- (0.6,-1.4);
			\draw (0.8,-1.4) -- (1.4,-1.4);
			\draw (1.6,-1.4) -- (2.2,-1.4);
			\draw (2.4,-1.4) -- (3.0,-1.4);
			\draw (3.2,-1.4) -- (3.8,-1.4);
			\draw (4.0,-1.4)
			.. controls (4.18,-1.4) and (4.42,0.6)
			.. (4.6,0.6);
			\draw (4.8,0.6)
			.. controls (4.98,0.6) and (5.22,-1.4)
			.. (5.4,-1.4);
			\draw (5.6,-1.4)
			.. controls (5.78,-1.4) and (6.02,0.6)
			.. (6.2,0.6);
			\draw (6.4,0.6)
			.. controls (6.58,0.6) and (6.82,-1.4)
			.. (7.0,-1.4);
			\draw (7.2,-1.4) -- (7.8,-1.4);
			\draw (8.0,-1.4) -- (8.6,-1.4);
			\draw (8.8,-1.4) -- (9.4,-1.4);
			\draw (9.6,-1.4) -- (10.2,-1.4);
			\draw (10.4,-1.4) -- (11.0,-1.4);
			
			\draw (0.6,0.6) -- (0.8,0.6);
			\draw (0.6,0.2) -- (0.8,0.2);
			\draw (0.6,-0.2) -- (0.8,-0.2);
			\draw (0.6,-0.6) -- (0.8,-0.6);
			\draw (0.6,-1.0) -- (0.8,-1.0);
			\draw (0.6,-1.4) -- (0.8,-1.4);
			
			\draw (1.4,0.6) -- (1.6,0.6);
			\draw (1.4,0.2) -- (1.6,0.2);
			\draw (1.4,-0.2) -- (1.6,-0.2);
			\draw (1.4,-0.6) -- (1.6,-0.6);
			\draw (1.4,-1.0) -- (1.6,-1.0);
			\draw (1.4,-1.4) -- (1.6,-1.4);
			
			\draw (2.2,-0.6) -- (2.4,-0.6);
			\draw (2.2,-1.0) -- (2.4,-1.0);
			\draw (2.2,-1.4) -- (2.4,-1.4);
			
			\draw (3.0,-0.2) -- (3.2,-0.2);
			\draw (3.0,-0.6) -- (3.2,-0.6);
			\draw (3.0,-1.0) -- (3.2,-1.0);
			\draw (3.0,-1.4) -- (3.2,-1.4);
			
			\draw (3.8,-1.4) -- (4.0,-1.4);
			
			\draw (4.6,-0.6) -- (4.8,-0.6);
			\draw (4.6,-1.0) -- (4.8,-1.0);
			\draw (4.6,-1.4) -- (4.8,-1.4);
			
			\draw (5.4,-1.4) -- (5.6,-1.4);
			
			\draw (6.2,-0.6) -- (6.4,-0.6);
			\draw (6.2,-1.0) -- (6.4,-1.0);
			\draw (6.2,-1.4) -- (6.4,-1.4);
			
			\draw (7.0,-1.4) -- (7.2,-1.4);
			
			\draw (7.8,-0.2) -- (8.0,-0.2);
			\draw (7.8,-0.6) -- (8.0,-0.6);
			\draw (7.8,-1.0) -- (8.0,-1.0);
			\draw (7.8,-1.4) -- (8.0,-1.4);
			
			\draw (8.6,-0.6) -- (8.8,-0.6);
			\draw (8.6,-1.0) -- (8.8,-1.0);
			\draw (8.6,-1.4) -- (8.8,-1.4);
			
			\draw (9.4,0.6) -- (9.6,0.6);
			\draw (9.4,0.2) -- (9.6,0.2);
			\draw (9.4,-0.2) -- (9.6,-0.2);
			\draw (9.4,-0.6) -- (9.6,-0.6);
			\draw (9.4,-1.0) -- (9.6,-1.0);
			\draw (9.4,-1.4) -- (9.6,-1.4);
			
			\draw (10.2,0.6) -- (10.4,0.6);
			\draw (10.2,0.2) -- (10.4,0.2);
			\draw (10.2,-0.2) -- (10.4,-0.2);
			\draw (10.2,-0.6) -- (10.4,-0.6);
			\draw (10.2,-1.0) -- (10.4,-1.0);
			\draw (10.2,-1.4) -- (10.4,-1.4);
			
			\draw[fill=white] (0.6,0.9) rectangle (0.8,1.5);
			
			\draw[fill=black] (1.4,0.9) rectangle (1.6,1.5);
			
			\draw[fill=white] (2.2,0.5) rectangle (2.4,1.5);
			\draw[fill=white] (2.2,-0.3) rectangle (2.4,0.3);
			
			\draw[fill=black] (3.0,0.9) rectangle (3.2,1.5);
			\draw[fill=black] (3.0,0.1) rectangle (3.2,0.7);
			
			\draw[fill=white] (3.8,-0.3) rectangle (4.0,1.5);
			\draw[fill=white] (3.8,-1.1) rectangle (4.0,-0.5);
			
			\draw[fill=black] (4.6,0.5) rectangle (4.8,1.5);
			\draw[fill=black] (4.6,-0.3) rectangle (4.8,0.3);
			
			\draw[fill=white] (5.4,-0.3) rectangle (5.6,1.5);
			\draw[fill=white] (5.4,-1.1) rectangle (5.6,-0.5);
			
			\draw[fill=black] (6.2,0.5) rectangle (6.4,1.5);
			\draw[fill=black] (6.2,-0.3) rectangle (6.4,0.3);
			
			\draw[fill=white] (7.0,-0.3) rectangle (7.2,1.5);
			\draw[fill=white] (7.0,-1.1) rectangle (7.2,-0.5);
			
			\draw[fill=black] (7.8,0.9) rectangle (8.0,1.5);
			\draw[fill=black] (7.8,0.1) rectangle (8.0,0.7);
			
			\draw[fill=white] (8.6,0.5) rectangle (8.8,1.5);
			\draw[fill=white] (8.6,-0.3) rectangle (8.8,0.3);
			
			\draw[fill=black] (9.4,0.9) rectangle (9.6,1.5);
			
			\draw[fill=white] (10.2,0.9) rectangle (10.4,1.5);
			\ ,
\end{scope}
}}

\title{Derivation of color projectors on multi-quark states using the Casimir operator}
 
\author[a]{St\'ephane Peign\'e,}

\author[b]{Cl\'ement Liesse}

\affiliation[a]{SUBATECH, Nantes Universit\'e, IMT Atlantique, IN2P3/CNRS, 4 rue Alfred Kastler, La Chantrerie BP 20722, 44307 Nantes, France}

\affiliation[b]{Laboratoire de Physique Subatomique et de Cosmologie,
Universit\'e Grenoble-Alpes, CNRS/IN2P3, Grenoble INP, 38000 Grenoble, France}

\emailAdd{stephane.peigne@subatech.in2p3.fr}
\emailAdd{clement.liesse@gmail.com}

\abstract{We present an elementary recursive construction of Hermitian projectors on multi-quark $\sun$ color states, based on the characterization of these states by their quadratic Casimir, and using the birdtrack pictorial technique.}

\keywords{quarks, $\sun$ irreducible representation, Casimir operator, birdtracks} 

\begin{document} 

\begin{flushright}
October 2026
\end{flushright}

\maketitle
\setcounter{footnote}{0}
\renewcommand{\thefootnote}{\arabic{footnote}}
\interfootnotelinepenalty=10000

\newpage

\section{Introduction}
\label{sec:intro}

In hadronic collisions at very high energies, realized for instance at the Large Hadron Collider, multi-parton systems can be produced in the early stage of the collision. The corresponding production cross sections are obtained from perturbative QCD calculations, which become increasingly complex as the number of partons increases.
One of the main challenges of multi-parton processes lies in their color structure, especially when the multi-parton state needs to be decomposed into its available color states, requiring knowledge of the Hermitian color projectors on these states. 

Over the past decade, the derivation of Hermitian color projectors has been the subject of several studies, all conducted using the {\it birdtrack} pictorial technique, a handy tool for dealing with the color aspects of QCD processes~\cite{Cvitanovic:1976am,Cvitanovic:2008zz,Dokshitzer:1995fv,Sjodahl:2015,Keppeler:2017kwt,Peigne:2023iwm,Peigne:2024srm}. In this context, the symmetry group of QCD is usually considered to be $\sun$, with the number $\Nc$ of quark colors taken as a general parameter ($\Nc =3$ corresponding to real-world QCD), and the ``color state'' of a system refers to the $\sun$ irreducible representation (irrep) to which that system belongs.
The algorithm developed in Ref.~\cite{Keppeler:2012ih} (and automated in Ref.~\cite{Chargeishvili:2024pnq}) to construct orthogonal color bases 
provides, in principle, access to the Hermitian projectors on the irreps of systems containing any number of partons. 
Systematic methods for obtaining Hermitian projectors starting from conventional Young projectors were developed in Ref.~\cite{Keppeler:2013yla}, and further explored in Refs.~\cite{Alcock-Zeilinger:2016xgf,Alcock-Zeilinger:2016cva,Alcock-Zeilinger:2016sxc}, where compact expressions for projectors were obtained using the MOLD algorithm~\cite{Alcock-Zeilinger:2016sxc}.

Although these algorithms seem fairly general, there are specific situations in which they cannot be directly applied. For example, for a given partonic system, one may be interested only in configurations where two subsystems are already projected on some $\sun$ irreps $R_1$ and $R_2$ (different from both the fundamental -- quark --  and adjoint -- gluon -- representations), and look for the Hermitian projectors on the irreps $\alpha_i$ appearing in the decomposition $R_1 \otimes R_2 = \sum_{i} \alpha_i\,$. In this case, the derivation of projectors can make use of methods other than the algorithms mentioned above.
For instance, it was shown in Ref.~\cite{Cougoulic:2025wpl} that the quadratic Casimir operator $\hat{C}$, acting on the $R_1 \otimes R_2$ {\it color space} (defined as the vector space of color maps $A: R_1 \otimes R_2 \to R_1 \otimes R_2$), proves to be very useful for simplifying the derivation of projectors.\footnote{Ref.~\cite{Cougoulic:2025wpl} considered the particular case $R_1=R_2=\Irrep{27}{}$, where $\Irrep{27}{}$ is the highest-dimensional $\sun$ irrep of a two-gluon system. But this case exhibits all features expected for any $R_1$ and $R_2$, and can thus be considered as generic.}
Indeed, the projector on an irrep $\alpha_i$ of quadratic Casimir $C_i$ is an eigenvector of $\hat{C}$ associated to the eigenvalue $C_i\,$. Thus, determining the eigenspaces of $\hat{C}$ is a very effective step in finding the projectors. The Casimir operator acts as a highly effective {\it filter} of the color space. We shall refer to the method of deriving projectors by first separating the color space into eigenspaces of $\hat{C}$ as ``Casimir operator filtering'' (COF).

In the situation considered in Ref.~\cite{Cougoulic:2025wpl}, the filtering is effective but not perfect, because some eigenspaces of $\hat{C}$ have a dimension $d>1\,$. 
This happens when several irreps $\alpha_i$ in the decomposition of $R_1 \otimes R_2$ have the same Casimir $C_i\,$. However, in that case it is easy to identify, in the eigenspace of $\hat{C}$ associated with the eigenvalue $C_i\,$, which eigenvector corresponds to the projector on a given irrep by using the irreps' symmetry properties w.r.t.~particle exchange~\cite{Cougoulic:2025wpl}. 
The COF method thus provides a fairly simple way to access the projectors on irreps that appear in the tensor product of any two irreps $R_1$ and $R_2\,$.

In the present study we use the COF method to construct the Hermitian projectors on the irreps of multi-quark systems in which the number of quarks is independent of $\Nc$, which is a much simpler problem than that considered in 
Ref.~\cite{Cougoulic:2025wpl}.\footnote{Ref.~\cite{Cougoulic:2025wpl} 
involves systems of quarks {\it and} antiquarks. 
Of course, by trading each antiquark for a (totally antisymmetric) system of $\Nc-1$ quarks, we may say that Ref.~\cite{Cougoulic:2025wpl} formally deals with multi-quark systems, but in which  the number of quarks is not independent of $\Nc$.} 
As is well-known, any $\sun$ irrep of such multi-quark systems can be obtained by starting from a single quark and piling up quarks one by one. At each step of this construction, we need to decompose a product of the type $R_1 \otimes q$, where $q$ denotes the fundamental (quark) representation. Since only inequivalent irreps (with distinct Casimirs) appear in this product, 
the Casimir operator filtering is optimal: the eigenspace of the operator $\hat{C}$ (applying to the $R_1 \otimes q$ color space) with eigenvalue $C_i$ has dimension $d=1$, and is spanned by the projector on the irrep of Casimir $C_i$ appearing in $R_1 \otimes q\,$. The COF method thus provides a nearly trivial way to obtain Hermitian projectors on multi-quark $\sun$ irreps. 

As already mentioned, Hermitian projectors of multi-quark systems can be obtained using known advanced algorithms~\cite{Keppeler:2012ih,Chargeishvili:2024pnq,Keppeler:2013yla,Alcock-Zeilinger:2016xgf,Alcock-Zeilinger:2016cva,Alcock-Zeilinger:2016sxc}.
However, the COF method presented here could also prove useful (not to mention the advantage of its extreme simplicity). 
The Hermitian projector on a multi-quark irrep $\al$ obtained with the COF method depends explicitly on the quadratic Casimirs of the successive irreps encountered in the building history of $\al$, rather than on their symmetry properties under quark permutations. 
This results in alternative birdtrack expressions for the projectors, that appear completely different (although they are mathematically identical) from those obtained using other algorithms (as \eg\ the MOLD algorithm~\cite{Alcock-Zeilinger:2016sxc}). 
In order to handle processes involving a large number of partons, it may be useful to have alternative expressions for the projectors, each of which offers its own advantages when used in symbolic computation programs.

The manuscript is organized as follows.
In section~\ref{sec:diquark}, we present all the steps of the COF method for the simplest system (namely, a pair of quarks), using basic tools of the birdtrack technique.
The method is generalized in section~\ref{sec:add-quark} to the case of any $\sun$ irrep combined with a single quark. 
Finally, in section~\ref{sec:illustration} the recursive construction of Hermitian projectors is illustrated using a concrete example, which can be generalized to express 
the Hermitian projector on any multi-quark state (defined by a Young tableau) in a compact algebraic form. 
The birdtrack conventions and basic rules used in this article are listed
in Appendix~\ref{app:rules}.
Readers in a hurry can skip directly to section~\ref{sec:illustration} to learn how to write, almost effortlessly, the birdtrack expression for any Hermitian multi-quark projector.

\section{The COF method applied to a quark pair}
\label{sec:diquark}

As is well-known from representation theory, an $\sun$ irrep can be characterized by its associated Young diagram (YD), encoding the irrep's symmetry properties under quark permutation. 
Knowing the YD of an irrep $R$, one can in principle obtain all $\sun$ invariants of the irrep, for instance its dimension (from the ``factor over hooks rule''~\cite{Georgi:1999wka}) and its quadratic Casimir $C_R\,$. The latter is given by 
(see \eg\ Ref.~\cite{Dietrich:2006cm}) 
\begin{equation}
\label{Casimir-formula}
C_R = T_F \left\{ b \, \Nc - \frac{b^2}{\Nc} + \sum_i r^2_i - \sum_i c^2_i \right\} \ ,
\end{equation}
where $T_F = \frac{1}{2}$ in our convention (see Appendix~\ref{app:rules} and eq.~\eq{eq:norm-conv}), $b$ is the YD's number of boxes, $r_i$ the number of boxes in its $i^{th}$ row, and $c_i$ the number of boxes in its $i^{th}$ column. 

A single quark $q^i$ ($i = 1\ldots \Nc$) spans the vector space 
(denoted by $q$) carrying the fundamental representation of $\sun$,
of Young diagram $\tikeq{\bs[scale=0.7]\tyng(0cm,-0.5mm,1)\es}$ 
and Casimir $\CF = \frac{\Nc^2-1}{2 \Nc}\,$.
To find the quark pair irreps, we combine two quarks according to the YD multiplication rules, 
\begin{equation}
\label{3times3-YD}
q \otimes q \ \sim \ 
\tikeq{\bs[scale=0.7]\tyng(0cm,0cm,1)\es} 
\otimes 
\tikeq{\bs[scale=0.7]\tyng(0cm,0cm,1)\es} 
\ = \ 
\tikeq{\bs[scale=0.7]\tyng(0cm,0cm,2)\es} 
\oplus 
\tikeq{\bs[scale=0.7]\tyng(0cm,0cm,1^2)\es} \ \ ,
\end{equation}
giving two ``diquark'' irreps, $s \equiv \tikeq{\bs[scale=0.6]\tyng(0cm,-0.5mm,2)\es}$ and 
$a \equiv \tikeq{\bs[scale=0.6]\tyng(0cm,-0.5mm,1^2)\es}\,\,$, which are respectively symmetric and antisymmetric under quark permutation. 

In birdtrack notation (see Appendix~\ref{app:rules}), 
the decomposition into irreps \eq{3times3-YD} means that the identity map $I$ of 
the diquark vector space $q \otimes q = q^{\otimes 2} \equiv q^2$ decomposes as: 
\begin{equation}
\label{comp-rel}
I \ \equiv \ \Idqq{1} \ = \ \Proj{s} + \Proj{a}\ \equiv \ \projblob{s}{1}  + \projblob{a}{1} \ = \  \sum_{\al = s, a} \projblob{\al}{1}  \ \ , 
\vspace{-2mm}
\end{equation}
where $\Proj{s}$ and $\Proj{a}$ are the mutually orthogonal Hermitian projectors (\ie, $\Proj{\al} \Proj{\beta} = \delta_{\al \beta} \Proj{\al}$ and $\Proj{\al}^\dagger = \Proj{\al}$ for $\al, \beta = s, a$) we are looking for. Knowing that the irreps can be characterized by their symmetry under quark permutation, $\Proj{s}$ and $\Proj{a}$ are given, respectively, by the symmetrizer and antisymmetrizer on the diquark color indices:
\be
\label{eq:PsPa-initial}
P_{s} \, = \,  \frac{1}{2} \big( \ \Idqq{1} + \Xqq{1}{0} \ \big) \, \equiv \, \PS{1} 
\ \  ; 
\ \ \ \ P_{a} \, = \, \frac{1}{2} \big( \ \Idqq{1} - \Xqq{1}{0} \ \big) \, \equiv \, \PA{1} 
\ \ . 
\ee

The goal of this section is to derive the well-known projectors \eq{eq:PsPa-initial} without relying on any knowledge of the symmetries of irreps, but by using the COF method, based on the characterization of diquark irreps by their quadratic Casimirs, which read (see eq.~\eq{Casimir-formula}) 
\begin{equation}
\label{eq:CsCa-values}
C_s = C_{_{\tikeq{\bs[scale=0.3]\tyng(0cm,-0.5mm,2)\es}}} \ = \ \Nc -\frac{2}{\Nc}+1 \ \ ; \ \ 
C_a = C_{^{\tikeq{\bs[scale=0.3]\tyng(0cm,-0.5mm,1^2)\es}}} \ = \ \Nc -\frac{2}{\Nc}-1 \ \ . 
\end{equation}
The advantage of the COF method 
is that it can be directly generalized to the case where a quark is combined with any irrep (see section~\ref{sec:add-quark}), allowing a straightforward recursive construction of the Hermitian projector on any multi-quark color state (see section~\ref{sec:illustration}). 

\vspace{2mm}
The derivation of $\Proj{s}$ and $\Proj{a}$ using the COF method relies on two observations:
\begin{itemize}
\item[(i)] The vector space of color maps $A$ from $q^2$ to $q^2$, called the ``$q^2$ color space'' and
denoted by ${\cal M}(q^2 \to q^2) \equiv {\cal M}$, 
has dimension 2, and $\{\Proj{s}, \Proj{a}\}$ is a basis of this space.
Indeed, any color map $A\in {\cal M}$ can be written as 
\begin{equation}
\label{eq:PsPaBasis}
\ampblob{A}{0.8} \, = \! \sum_{\al, \beta = s, a} \!\! \projblob{\al}{0.8} \hskip -0.8mm
\Mblob{A}{0.8}  \hskip -0.5mm \projblob{\beta}{0.8} = 
\sum_{\al} \, \projblob{\al}{0.8} \hskip -0.8mm
\Mblob{A}{0.8}  \hskip -0.5mm \projblob{\al}{0.8}
= \sum_{\al} \, a_{\al}  \Proj{\al} \ , 
\end{equation}
where we first inserted the completeness relation \eq{comp-rel} on the left and right, then used Schur's lemma which states that non-equivalent irreps (\ie, having different YD's) cannot ``map to each other'',  implying $\beta = \al$, and finally used 
$\Proj{\al} A \Proj{\al} \propto \Proj{\al}$, which also follows from Schur's lemma~\cite{Georgi:1999wka}.
\item[(ii)] Defining pictorially the quadratic Casimir operator $\hat{C}$ acting on ${\cal M}$
as~\cite{Peigne:2024srm,Cougoulic:2025wpl}
\begin{equation}
\label{eq:Cdef}
\hat{C} \cdot \ampblob{A}{1} \ = \  \Casimirop{A}{1} \ \ ,
\end{equation}
where the grey ellipse represents the sum of gluon attachments to the two quark lines (see eq.~\eq{eq:color-cons}), we easily see that $\Proj{\al}$ ($\al = s, a$) is an eigenvector of $\hat{C}$ for the eigenvalue $C_{\al}\,$. Indeed, we can write
\begin{equation}
\hat{C} \cdot \Proj{\al} \ = \ \hat{C} \cdot \Proj{\al}^2 \ = \ \CasPPinside{1} \ = \ \CasPPoutside{1}  
\ = \ C_{\al} \Proj{\al} \ . 
\end{equation}
Here we used color conservation,\footnote{Color conservation implies that a grey ellipse drawn at a given intermediate state of a color map can be translated to any other intermediate state, see eq.~\eq{eq:color-cons} in Appendix~\ref{app:rules}.} and the last equality follows from Schur's lemma and the fact that in the subspace ${\rm img}(\Proj{\al})$, the quadratic Casimir $C_{\al}$ is by definition the proportionality coefficient between $T^{a}_{\al} T^{a}_{\al}$ (with $T^{a}_{\al}$ the $\sun$ generators in irrep $\al$) and the identity operator in the irrep $\al$, namely $\unit_{\al} = \Proj{\al}\,$. 
\end{itemize}
Therefore, in the basis $\{\Proj{s}, \Proj{a}\}$ of ${\cal M}$, the operator $\hat{C}$ and the operators $\hat{P}_s$ and $\hat{P}_a$ projecting on the $\hat{C}$ eigenspaces simply read 
\begin{equation}
\label{eq:Cdiag}
\hat{C} \, = \, 
\left( \begin{matrix}  
C_s \ & \ 0 \  \\[1mm] 
0 \ & \ C_a 
\end{matrix}  \right)  \ ; \ \ 
\hat{P}_s \, = \, 
\left( \begin{matrix}  
1 \ & \ 0 \  \\[1mm] 
0 \ & \ 0 
\end{matrix}  \right) 
\ ; \ \ 
\hat{P}_a \, = \, 
\left( \begin{matrix}  
0 \ & \ 0 \  \\[1mm] 
0 \ & \ 1 
\end{matrix}  \right) \ ; 
\ \ \ \ \hat{P}_s + \hat{P}_a \, = \,  \left( \begin{matrix}  1 \ & \ 0 \  \\[1mm] 0 \ & \ 1 \end{matrix}  \right) \, \equiv \,  \hat{I}  \ . 
\end{equation}
The action of $\hat{P}_{\al}$ on ${\cal M}$ is nothing more than the multiplication by $\Proj{\al}\,$, to the left or equivalently to the right, as follows directly from~\eq{eq:PsPaBasis}:
\begin{equation}
\label{eq:alpha-mult}
\hat{P}_{\al} \cdot A \,=\, \Proj{\al} \, A \,=\,  A \, \Proj{\al} \ . 
\end{equation}
Similarly, using \eq{eq:Cdef} and color conservation, we see that the action of $\hat{C}$ 
on $A \in {\cal M}$ is $\hat{C} \cdot A =  C A = A C$, where the color map $C$ is defined by
\begin{equation}
\label{eq:Cmap}
C \ \equiv \ \CasCal{1} \ = \ \hat{C} \cdot I \ \ . 
\end{equation}

Basis-independent expressions of $\hat{P}_{\al}$ directly follow from \eq{eq:Cdiag}, 
\begin{equation}
\label{eq-PsPa-Frobenius}
\hat{P}_s \, = \, \frac{\hat{C} - C_a \, \hat{I}}{C_s - C_a} \ \ ; \ \ \ \ 
\hat{P}_a \, = \, \frac{\hat{C} - C_s \, \hat{I}}{C_a - C_s} \ .
\vspace{2mm}
\end{equation}
These expressions are explicitly $\sun$ invariant, since they only depend on Casimirs and on the operator $\hat{C}$ defined in invariant form by \eq{eq:Cdef}. 
Using \eq{eq:alpha-mult} for $A = I$, we get 
\be
\label{eq:PsPa-Casimir}
\Proj{s} \, = \hat{P}_s \cdot I \,=\, \frac{C - C_a \, I}{C_s - C_a} \ \ ; \ \ \ \ 
\Proj{a} \, = \hat{P}_a \cdot I \,=\, \frac{C  - C_s \, I}{C_a - C_s}  \ \ . 
\ee

The projectors \eq{eq:PsPa-Casimir} obtained using the COF method are mutually orthogonal by construction. Indeed, using \eq{eq:alpha-mult} and the fact that the operators $\hat{P}_{\al}$ 
(acting on ${\cal M}$) are themselves such projectors, one can write 
$\Proj{\al}^2 = \hat{P}_\al^2 \cdot I = \hat{P}_\al \cdot I =\Proj{\al}$ (for $\al = s, a$), and 
$\Proj{s} \Proj{a} = \hat{P}_s \hat{P}_a \cdot I = 0\,$. The projectors 
\eq{eq:PsPa-Casimir} are also Hermitian (since $I$ and $C$ are),
and obviously satisfy the completeness relation \eq{comp-rel}. 
Using now the Fierz identity (see Appendix~\ref{app:rules} and eq.~\eq{eq:Fierz}), the color map \eq{eq:Cmap} can be written as 
\begin{equation}
\label{eq:Cmap-2}
C \, = \, \CasCal{1} \,=\, 2 \CF  \, \Idqq{1} -2\ \Vqq{1}{0}  
\,=\, \Big( 2 \CF -\frac{1}{\Nc} \Big) \,  \Idqq{1} \, + \, \Xqq{1}{0} \ \ .
\vspace{2mm}
\end{equation}
We can then immediately verify that $\Proj{s}$ and $\Proj{a}$ given by \eq{eq:PsPa-Casimir}
coincide with the projectors \eq{eq:PsPa-initial} (using the values of $C_s$ and $C_a$ given by~\eq{eq:CsCa-values}).  

Instead of the operator $\hat{C}$ used to filter the color space, we could have chosen any operator $\hat{C} - \lambda \hat{I}$, which also has $\Proj{\al}$ as eigenvectors, with associated eigenvalues $C_\al - \lambda\,$. The obtained projectors would then be written as in \eq{eq:PsPa-Casimir}, up to the replacements $C \to C - \lambda I$ and $C_\al \to C_\al - \lambda$ (which obviously gives mathematically identical expressions). 
For instance, choosing the ``interaction potential'' operator $\hat{V} = \hat{C} - 2 \CF \hat{I}$, whose action on ${\cal M}$ is the multiplication (to the left or right) by the color map
\begin{equation}
V \,=\, C - 2\CF \, I \,=\, - 2\ \Vqq{1}{1} \ \ , 
\end{equation}
the projectors \eq{eq:PsPa-Casimir} are written as
\begin{equation}
\label{eq:PsPa-potential}
\Proj{s} \,=\, \frac{V - v_a \, I}{v_s  - v_a} \ \ ; \ \ \ \ 
\Proj{a} \,=\, \frac{V - v_s \, I}{v_a - v_s}  \ \ , 
\end{equation}
where $v_\al = C_\al - 2 \CF\,$. 
Another convenient choice is to use the ``exchange operator''
$\hat{X}= \hat{C} - \big( 2 \CF -\frac{1}{\Nc} \big) \hat{I}\,$, which
acts on ${\cal M}$ as the multiplication 
by the color map
\begin{equation}
\label{eq:Xmap}
X \,=\,  C - \Big( 2 \CF -\frac{1}{\Nc} \Big) \, I \,=\, \Xqq{0.9}{1} \ \ . 
\end{equation}
With this choice, the projectors \eq{eq:PsPa-Casimir} read
\begin{equation}
\label{eq:PsPa-exchange}
\Proj{s} \,=\, \frac{X - x_a \, I}{x_s  - x_a} \ \ ; \ \ \ \ 
\Proj{a} \,=\, \frac{X - x_s \, I}{x_a - x_s}  \ \ ,
\end{equation}
where $x_\al = C_\al - 2 \CF + \frac{1}{\Nc} = C_\al - \Nc + \frac{2}{\Nc}\,$. Using \eq{eq:CsCa-values}, we get $x_s = 1$ and $x_a = -1$ (which is not surprising if we recall that the irreps $s$ and $a$ are symmetric and antisymmetric under permutation), and we recover once again the projectors \eq{eq:PsPa-initial}. 

By using only the Casimir operator $\hat{C}$ (or, depending on our preference, the potential operator $\hat{V}$ or the exchange operator $\hat{X}$), we were able to derive the projectors \eq{eq:PsPa-initial} on diquark color states.
If the goal were simply to obtain such simple results, the method might seem a bit tedious. However, as mentioned earlier, the interest of the COF method lies in the fact that it generalizes almost trivially to systems containing any number of quarks, as we will see in the following sections.

\section{Adding a quark to a multi-quark color state} 
\label{sec:add-quark}

The COF method used in the previous section to find the projectors on diquark irreps can be directly applied to the tensor product of any $\sun$ irrep with a single quark. 

Suppose we have already projected a system of $b-1$ quarks on a certain $\sun$ irrep $\al$, using the Hermitian projector $\Proj{\al}$ (belonging to the space ${\cal M}^{(b-1)}$ of color maps $A: q^{(b-1)} \to q^{(b-1)}$, where $q^{(b-1)} \equiv q^{\otimes (b-1)}$) represented pictorially as
\begin{equation}
\Proj{\al} \ = \projblobmulti{\al}{1.1} \ \ .
\end{equation}

The tensor product of the irrep $\al$ with a single quark decomposes as 
$\al \otimes q = \mathop{\oplus}_{i=1}^{n} \beta_i\,$, where the $n$ irreps $\beta_i$
have distinct quadratic Casimirs $C_i$ and are therefore not equivalent.\footnote{The non-equivalence of the $\beta_i$'s follows from the simplicity of the YD multiplication rules when one factor is a single quark. For illustration, we show below the tensor product of some irrep $\al$ (for $b-1= 7$) with one quark (numbered by $b=8$), decomposing into three distinct YD's, \ie\ three inequivalent irreps:
\vspace{-2mm} 
\begin{equation}
\tikeq{\bs[scale=0.6]\tyng(0cm,0cm,5,2)\es} 
\otimes \, \tikeq{\bs[scale=0.6]\tyoung(0cm,-0.5mm,8)\es} 
\ = \ 
\tikeq{\bs[scale=0.6]\tyoung(0cm,0cm,\ \ \ \ \ 8,\ \ )\es} \,\oplus\,
\tikeq{\bs[scale=0.6]\tyoung(0cm,0cm,\ \ \ \ \ ,\ \ 8)\es} \,\oplus \,
\tikeq{\bs[scale=0.6]\tyoung(0cm,0cm,\ \ \ \ \ ,\ \ ,8)\es} \nn \ \ .
\vspace{-2mm} 
\end{equation}
It is also easy to see that the Casimirs $C_i$ are different. 
Anticipating a bit, we have $C_i = C_\al + \CF - \frac{b-1}{\Nc} + x_i$ (see~\eq{eq:Xi-formula}), with $x_i = r(b) - c(b)$ (see~\eq{eq:Xi-handy}), where $r(b)$ and $c(b)$ are defined below eq.~\eq{eq:Xi-handy}. Clearly, $x_i$, and therefore $C_i$, is different for different $\beta_i$'s. (For instance, in the above example, the three irreps listed from left to right have a decreasing value of $r(b)-c(b)$, and thus a decreasing $C_i\,$.)}
The Hermitian projectors $P_i$ on $\beta_i$ have the general form 
\begin{equation}
\label{eq:beta-i}
P_i \ = \ \projmultalbeta{\al}{B_i}{1} \ \ , 
\end{equation}
thus being maps of
$\left[q^{(b-1)} \right]_{\al} \!\otimes q \to \left[q^{(b-1)} \right]_{\al} \! \otimes q$ (the subscript $\al$ indicating that the system of the first $b-1$ quarks is in irrep $\al$), 
the set of which will be denoted as ${\cal M}_{\al}^{(b)}$. The $P_i$'s must also satisfy the completeness relation:
\vspace{-2mm}
\begin{equation}
\label{mult-comp}
\sum_{i=1}^{n} P_i \ = \ \Idmult{\al}{1} \ \equiv \ I \ \ ,
\end{equation}
where $I$ is the identity map of ${\cal M}_{\al}^{(b)}$. Our goal here is to derive 
the $\sun$ invariant, birdtrack expressions of the blobs $B_i$ 
to be inserted into \eq{eq:beta-i}, such that the $P_i$'s form a complete set of mutually orthogonal Hermitian projectors.

\vspace{2mm}
As in the case of a quark pair discussed in section~\ref{sec:diquark}, it can easily be shown that:
\begin{itemize}
\item[(i)] The set of projectors $P_i$ is a basis of ${\cal M}_{\al}^{(b)}$; 
\item[(ii)] The quadratic Casimir operator $\hat{C}: {\cal M}_{\al}^{(b)} \to {\cal M}_{\al}^{(b)}$ (defined for general $b$ similarly to the diquark case $b=2$, see eq.~\eq{eq:Cdef}) is diagonal in this basis, $\hat{C} \cdot P_i = C_{i} \, P_i\,$. 
\vspace{2mm}
\end{itemize} 
Therefore, the operator $\hat{C}$ and its eigenspaces can be used to filter the color space ${\cal M}_{\al}^{(b)}$ and obtain the projectors $P_i\,$. But we might as well use the interaction potential operator, now defined as $\hat{V} \equiv \hat{C} - (C_\al + \CF ) \, \hat{I}$, or the exchange operator 
\begin{equation}
\label{X-op}
\hat{X} \ \equiv \ \hat{C} - \Big( C_\al + \CF -\frac{b-1}{\Nc} \Big) \, \hat{I} \ \ ,
\end{equation}
where $\hat{I}$ is the identity operator acting on ${\cal M}_{\al}^{(b)}$, 
corresponding to multiplication by the map $I$ defined in eq.~\eq{mult-comp}. 
In what follows we will choose the operator $\hat{X}$. 

Thus, in the basis $\{ P_i \}$, $\hat{X}$ is diagonal, 
with $n$ distinct eigenvalues given by 
\begin{equation}
\label{eq:Xi-formula}
x_i \,=\,  C_i - C_\al - \CF + \frac{b-1}{\Nc} \ . 
\end{equation}
Note that $x_i$ can be regarded as an ``exchange potential'' between the two parent irreps $\al$ and $q$, when they are combined in the color state $\beta_i\,$.

The operator $\hat{P}_i$ projecting a color map of ${\cal M}_{\al}^{(b)}$ on its component $\propto P_i$ then reads, in invariant form 
\begin{equation}
\label{eq-Pbeta-Frobenius}
\hat{P}_i \ = \ \bigg(\, \mathop{\prod}_{k\neq i} \, \frac{\hat{X} - x_{k} \, \hat{I}}{x_{i} - x_{k}}  \, \bigg) \cdot \hat{I} 
\ \ ,
\end{equation}
where the product runs over $k=1 \ldots n$ with the value $k=i$ being omitted, and the factor $\hat{I}$ on the right is a reminder that $\hat{X}$ and $\hat{P}_i$ are defined as operators acting on ${\cal M}_{\al}^{(b)}$. Equation \eq{eq-Pbeta-Frobenius} is a generalization of \eq{eq-PsPa-Frobenius} to cases where $n > 2$. 

Using the pictorial expression of $\hat{C}$, color conservation, and the Fierz identity~\eq{eq:Fierz},
we easily verify that the action of $\hat{X}$ defined in eq.~\eq{X-op} 
corresponds to multiplication by
\begin{equation}
\label{eq:Xop-bird}
X \ = \ \sum_{k=1}^{b-1} \ \Xmult{1.2} \ \ ,
\end{equation}
which generalizes \eq{eq:Xmap} to any $b \geq 2$, and justifies calling $\hat{X}$ the ``exchange operator''.

Since the action of $\hat{P}_i\,$, $\hat{X}$ and $\hat{I}$ on ${\cal M}_{\al}^{(b)}$ amounts to multiplying by $P_i\,$, $X$ and $I$, we can ``remove the hats'' in eq.~\eq{eq-Pbeta-Frobenius}  (as we did in \eq{eq-PsPa-Frobenius} to obtain \eq{eq:PsPa-Casimir}) and write
\begin{equation}
\label{eq:Pbeta-i-X}
P_i \ = \hat{P}_i \cdot I \ =\ 
I \, \bigg( \mathop{\prod}_{k\neq i} \, \frac{X - x_{k} \, I}{x_{i} - x_{k}} \, \bigg) \, I  
\ \ ,
\end{equation}
which generalizes eq.~\eq{eq:PsPa-exchange}. In eq.~\eq{eq:Pbeta-i-X}, 
the factor $I$ on the right reminds us that $X$ defined by \eq{eq:Xop-bird} should be multiplied only by maps of ${\cal M}_{\al}^{(b)}$. Since $I = I^2$ and $X$ commutes with $I$ (because $\hat{C}$ commutes with $\hat{I}$, from color conservation), we can add a factor $I$ to the left in order to put $P_i$ in the form \eq{eq:beta-i} (and thus infer the expression of the blob $B_i$). But this factor $I$ is superfluous. Similarly, the factor $I$ in $X - x_{k} \, I$ being always multiplied by itself on the right, it can be replaced by the identity map of the vector space $q^b$ 
and can therefore be safely omitted. 

The projector $P_i$ on the irrep $\beta_i$ can thus be written as follows (where the exact order of the two blobs in the r.h.s.~is irrelevant)
\begin{equation}
\label{proj-beta-i}
P_i \ = \ 
\bigg( \mathop{\prod}_{k\neq i} \, \frac{X - x_{k}}{x_{i} - x_{k}} \, \bigg) \, I  
\ = \ 
\projmultalbetacut{\al}{\underset{k\neq i}{\prod} \frac{X - x_{k}}{x_{i} - x_{k}}}{1.3} \ \ .
\end{equation}

In summary, the expression \eq {proj-beta-i} should be transparent. When adding a quark to a system of $b-1$ quarks in irrep $\al$, we obtain $n$ inequivalent irreps $\beta_i\,$, which can be distinguished by their eigenvalue $x_i$ (given by \eq{eq:Xi-formula}) w.r.t.~the exchange operator $\hat{X}$ between the $b-1$ quarks and the added quark. (The action of $\hat{X}$ corresponds to the multiplication by the map $X$ defined in \eq{eq:Xop-bird}). 
The role of the blob shown on the right of eq.~\eq{proj-beta-i} (which directly follows from eq.~\eq{eq-Pbeta-Frobenius}) is to project on the irrep of eigenvalue $x_i\,$, by removing the unwanted irreps of eigenvalues $x_k$ with $k \neq i\,$.

By construction, the $n$ projectors $P_i$ are mutually orthogonal, Hermitian, satisfy the completeness relation \eq{mult-comp}, and are eigenvectors of $\hat{C}$ for the eigenvalues $C_i\,$. 
So these are the projectors on the irreps $\beta_i$ appearing in $\al \otimes q = \mathop{\oplus}_{i=1}^{n} \beta_i$ that we were looking for. 

\section{Using the COF method recursively} 
\label{sec:illustration}

By using eq.~\eq{proj-beta-i} iteratively, writing the birdtrack expression of the Hermitian projector on an $\sun$ irrep is almost self-evident. 

The method can be simply illustrated by taking the example of the irrep with a construction history given by the Young tableau:\footnote{We recall that in the decomposition of a system of $b$ quarks with $b>2$, the same $\sun$ irrep (\ie, the same YD) may have several occurrences. This happens when the $b$ boxes of the irrep's YD can be filled with the numbers $1 \ldots b$ in different ways, called Young tableaux. A Young tableau specifies the ``construction history'' of the irrep, \ie, how the $b$ quarks are successively combined to build the irrep's YD. See \eg\ Ref.~\cite{Alcock-Zeilinger:2016sxc}.}
\begin{equation}
\label{YT-example}
\tikeq{\bs[scale=0.8]\tyoung(0cm,0cm,12467,35,8)\es}  \ \ . 
\end{equation}
At each stage of the construction, when the first $b-1$ quarks have already been projected on the irrep $\al$ of Young tableau given by the first $b-1$ boxes, and the corresponding Hermitian projector has already been written down, we simply need to know the accessible irreps $\beta_k$ resulting from adding one quark, and the eigenvalues $x_k$ distinguishing these irreps. The eigenvalues are given by \eq{eq:Xi-formula}. Using~\eq{Casimir-formula} to find $C_\al$ and $C_k$ from the YD's of the irreps $\al$ and $\beta_k$ (which have $b-1$ and $b$ boxes, respectively), it turns out that \eq{eq:Xi-formula} can be simplified into the handy expression 
\begin{equation}
\label{eq:Xi-handy}
x_k = r(b) - c(b)  \ ,
\end{equation}
where $r(b)$ (resp.~$c(b)$) is the length 
of the row (resp.~column), in the YD of ${\beta_k}\,$, 
occupied by the last quark added (\ie, quark number $b$). 

Starting from a single quark and adding one quark, the two resulting irreps 
$\beta_k$ ($k=1,2$) are given by (see eq.~\eq{3times3-YD}) 
\vspace{-4mm}
\begin{equation}
\label{3times3-YT}
\tikeq{\bs[scale=0.7]\tyoung(0cm,0cm,1)\es}
\otimes 
\tikeq{\bs[scale=0.7]\tyoung(0cm,0cm,2)\es} 
\ = \ \beta_1 \oplus \beta_2 
\ = \ 
\stackrel{1}{\tikeq{\bs[scale=0.7]\tyoung(0cm,0cm,12)\es}}
\oplus
\stackrel{-1}{\tikeq{\bs[scale=0.7]\tyoung(0cm,0cm,1,2)\es}} \ \ , 
\end{equation}
where now the building history of each irrep is specified by its Young tableau, above which we indicated the corresponding irrep's eigenvalue $x_k\,$, obtained with the rule 
\eq{eq:Xi-handy}. Knowing the $x_k$'s, we can use \eq{proj-beta-i} to write the projector on  
$\tikeq{\bs[scale=0.7]\tyoung(0cm,0.5mm,12)\es}$ as
\begin{equation}
P\left( \tikeq{\bs[scale=0.7]\tyoung(0cm,0cm,12)\es} \right)
\ = \ \projblob{\mathsize{8}{\frac{X - x_{2}}{x_{1} - x_{2}}}}{1.6} 
\ = \ \projblob{\mathsize{8}{\frac{X +1}{2}}}{1.6}  \ \ , 
\end{equation}
where $X$ is given by \eq{eq:Xop-bird} for $b=2$, \ie\ $X=\Xqq{0.8}{0}$ (see \eq{eq:Xmap}). 
We obviously recover the projector $P_{s}$ on the symmetric diquark state (see eq.~\eq{eq:PsPa-initial}). 

To continue building the projector of the irrep 
\eq{YT-example}, we need the projector on 
$\tikeqbis{\bs[scale=0.7]\tyoung(0cm,2mm,12,3)\es}$, the projector on 
$\tikeq{\bs[scale=0.7]\tyoung(0cm,0cm,12)\es}$ being known. We thus need the irreps appearing in $\tikeq{\bs[scale=0.7]\tyoung(0cm,0cm,12)\es} \otimes 
\tikeq{\bs[scale=0.7]\tyoung(0cm,0cm,3)\es}$ and their corresponding eigenvalues $x_k\,$: 
\vspace{-1mm}
\begin{equation}
\label{qqq-YT}
\tikeq{\bs[scale=0.7]\tyoung(0cm,0cm,12)\es} \otimes 
\tikeq{\bs[scale=0.7]\tyoung(0cm,0cm,3)\es} 
\ = \ \beta_1 \oplus \beta_2 
\ \equiv \ \ 
\stackrel{2}{\tikeq{\bs[scale=0.7]\tyoung(0cm,0cm,123)\es}} \oplus
\stackrel{-1}{\tikeq{\bs[scale=0.7]\tyoung(0cm,0cm,12,3)\es}}
 \ \ . 
\end{equation}
Applying now \eq{proj-beta-i} we get
\vspace{-1mm}
\begin{equation}
\label{P123}
P \Big( \tikeq{\bs[scale=0.7]\tyoung(0cm,0cm,12,3)\es} \Big) \ = \ 
\projthree{\tikeq{\bs[scale=0.5]\tyoung(0cm,0cm,12)\es}}{\frac{X - 2}{-1 - 2}}{1.4}{1.3} 
 \ = \ 
\projthree{\frac{X +1}{2}}{\frac{X-2}{-3}}{1.4}{1.3} 
\ \ .
\end{equation} 
Here, in the rightmost blob, $X$ is given by \eq{eq:Xop-bird} for $b=3$ (it contains two terms). 
In general, the map $X$ appearing in some blob is defined by \eq{eq:Xop-bird} for the value of $b$ given by the number of lines entering that blob, and thus contains $b-1$ terms.

To obtain the projector on $\tikeq{\bs[scale=0.7]\tyoung(0cm,0cm,124,3)\es}$, we need: 
\vspace{-2mm}
\begin{equation}
\label{123times4}
\tikeq{\bs[scale=0.7]\tyoung(0cm,0cm,12,3)\es} \otimes 
\tikeq{\bs[scale=0.7]\tyoung(0cm,0cm,4)\es} 
\ = \ \beta_1 \oplus \beta_2 \oplus \beta_3
\ \equiv \ \ 
\stackrel{2}{\tikeq{\bs[scale=0.7]\tyoung(0cm,0cm,124,3)\es}} \oplus
\stackrel{0}{\tikeq{\bs[scale=0.7]\tyoung(0cm,0cm,12,34)\es}} \oplus
\stackrel{-2}{\tikeq{\bs[scale=0.7]\tyoung(0cm,0cm,12,3,4)\es}} 
\ \ .
\end{equation}
Using \eq{proj-beta-i} for $\al = \tikeq{\bs[scale=0.7]\tyoung(0cm,0cm,12,3)\es}$
and plugging \eq{P123} we obtain
\vspace{-2mm}
\begin{equation}
P \Big( \tikeq{\bs[scale=0.7]\tyoung(0cm,0cm,124,3)\es} \Big)
\ = \ \projfour{\frac{X +1}{2}}{\frac{X-2}{-3}}{\frac{X(X+2)}{8}}{1.7}{1.3} \ \ ,
\end{equation} 
where in the rightmost blob, $X$ is given by \eq{eq:Xop-bird} for $b=4$. Note that this blob contains two factors ($X$ and $X+2$) 
because there are 3 irreps in the decomposition \eq{123times4}. 

And so on... After a little practice, all of the steps above can be performed visually, simply by observing  the Young tableau \eq{YT-example}, and using the simple rule \eq{eq:Xi-handy}. 
We'll leave it to the reader to verify in a minute that the projector on this irrep is written as: 
\begin{equation}
\label{Proj8}
P\bigg(\, \tikeq{\bs[scale=0.7]\tyoung(0cm,0cm,12467,35,8)\es} \, \bigg) \ = \ 
\projeight{X^{-1}_{1}}{X^{2}_{-1}}{X^{0}_{2} X^{-2}_{2}}{X^{3}_{0} X^{-2}_{0}}{X^{1}_{3} X^{-2}_{3}}{X^{1}_{4} X^{-2}_{4}}{X^{5}_{_{-2}} X^{1}_{_{-2}}}{1.6}{1} \ \ ,
\end{equation}
where we used the 
notation $X^{x_k}_{x_i} \equiv \frac{X- x_k}{x_i - x_k}$. 
Again, we stress that the 
form of the map 
$X$ 
in each blob depends on the number $b$ of lines entering that blob according to eq.~\eq{eq:Xop-bird}.  

Let's conclude by a few general remarks. 

First, let us note that the COF method provides an intuitive way of seeing that projectors on equivalent irreps, although associated with the same Young diagram, are orthogonal to each other. Indeed, the construction histories (\ie, Young tableaux) of two equivalent irreps necessarily differ at some point. At the stage of construction where they differ, the two corresponding projectors are orthogonal, since they project their common ancestor space onto two orthogonal subspaces. These two orthogonal projectors 
are eventually embedded in the construction histories of the two equivalent irreps, but using the commutativity of the blobs in eq.~\eqref{proj-beta-i}, they can always be placed side by side in the product of the two final projectors. The product of projectors on the equivalent irreps therefore vanishes. 

As mentioned in the Introduction, several algorithms have been developed to determine Hermitian projectors on $\sun$ irreps~\cite{Keppeler:2012ih,Chargeishvili:2024pnq,Keppeler:2013yla,Alcock-Zeilinger:2016xgf,Alcock-Zeilinger:2016cva,Alcock-Zeilinger:2016sxc}. These algorithms are based on the symmetry properties of the irreps, whereas the COF method uses their quadratic Casimir. This results in birdtrack expressions of the projectors that look completely different depending on the algorithms used. 
As an illustration, using the MOLD algorithm~\cite{Alcock-Zeilinger:2016sxc}, one finds that the projector on the irrep of Young tableau \eqref{YT-example} can be written as: 
\begin{equation}
	\label{Pmold}
	P\bigg(\, \tikeq{\bs[scale=0.7]\tyoung(0cm,0cm,12467,35,8)\es} \, \bigg) \ = \ \beta \ 
	\pMOLDeight{0.9}{1} \ \ ,
\end{equation}
where a white rectangle (resp.~black rectangle) denotes a symmetrizer  (resp.~antisymmetrizer) over the quark lines entering that rectangle, and 
the normalization factor $\beta$ 
can be determined either from the idempotence of the projector or 
by requiring that its trace be equal to the 
dimension of the corresponding irrep, which yields 
$\beta = \frac{32768}{147}$. 

We have explicitly checked that the expressions \eqref{Proj8} and \eqref{Pmold} 
are mathematically identical. When expanding these expressions, the number of terms obtained is huge, namely, about $2 \times 10^{8}$ 
for COF and $7 \times 10^{13}$ 
for MOLD.\footnote{Of course, in each case, the expression simplifies considerably after expansion, since the projector must be a linear combination of the 
$8!$ 
permutations of the system of $8$ quarks.}
As a consequence, in our implementation using Python, the COF expression is obtained significantly faster than the MOLD one. This difference can become important for systems containing a large number of quarks, for which the computation of the projector becomes increasingly demanding. On the other hand, the MOLD expression makes the symmetry properties of the irrep 
visible through the symmetrizers and antisymmetrizers.

Finally, the recursive COF method, based on~\eq{proj-beta-i} and illustrated by the example~\eq{Proj8}, can be presented in the form of a concise algebraic formula, 
giving the Hermitian color projector corresponding to a Young tableau directly from a simple encoding of this tableau. We characterize a Young tableau $Y_{\rm T}$  
containing $b$ boxes
by a list of integers \(l_i\) ($i=1 \ldots b$), 
where \(l_i\) is the number of the row containing the $i^{\rm th}$ quark. 
(For instance, the Young tableau \eq{YT-example} is defined by the list
$(1, 1, 2, 1, 2, 1, 1, 3)$.)
Let \(N_m^{(i)}\) denote the number of boxes in the $m^{\rm th}$ row at step $(i)$
(defined as the moment of the construction history when the first \(i\) quarks have been added), and 
$\mathcal{A}^{(i)}$  be the set of rows accessible to the $i^{\rm th}$ box 
at step $(i-1)\,$:
\vspace{-2mm}
\begin{equation}
N_m^{(i)} = \sum_{k=1}^{i} \delta_{l_k , \, m} \ , \quad 
\mathcal{A}^{(i)} 
= \{ j\geq 1 \ | \  N_j^{(i-1)} < N_{j-1}^{(i-1)} \}  \quad \ \ (N_0^{(i-1)} \equiv + \infty ) \, .
\vspace{-2mm}
\end{equation}
Using the above notations and eq.~\eq{eq:Xi-handy}, 
the eigenvalue 
$x_{l_i}^{(i)}$ 
corresponding to the position taken by the 
$i^{\rm th}$ box at step $(i)$, 
and the eigenvalues 
$x_j^{(i)}$ 
associated with the other rows 
of $\mathcal A^{(i)}$ (\ie\ with $j \neq l_i$), 
are given by 
\begin{equation}
\label{eq:kiyi}
x_{l_i}^{(i)} = N_{l_i}^{(i)} - l_i \, , \qquad 
x_j^{(i)} = N_j^{(i-1)} + 1 - j \  .
\end{equation}
The projector on the irrep corresponding to the tableau $Y_{\rm T}$ then reads
\begin{equation}
\label{eq:alg-exp}
P(\,Y_{\rm T}\,) = \prod_{i=2}^b \  \prod_{\substack{j \in \mathcal{A}^{(i)} \\ j \neq l_i}} \ \frac{X^{(i)} - x_j^{(i)}}{x_{l_i}^{(i)} - x_j^{(i)}} \ , \qquad X^{(i)} = \sum_{k=1}^{i-1} (ki) \ ,
\end{equation}
where $X^{(i)}$ is the map defined by eq.~\eq{eq:Xop-bird} (for $b=i$), 
and \( (ki) \) denotes the transposition of \(k\) and \(i\). 
The equations above provide a simple way to automate the calculation of the Hermitian projector associated with any Young tableau encoded by the list of integers \(l_i\)\,.

\appendix

\section{Basic pictorial rules}
\label{app:rules}

The birdtrack technique~\cite{Cvitanovic:1976am,Cvitanovic:2008zz,Dokshitzer:1995fv,Sjodahl:2015,Keppeler:2017kwt,Peigne:2023iwm,Peigne:2024srm} involves drawing color graphs, called birdtracks, instead of writing mathematical symbols with color indices. The basic pictorial rules slightly vary in the literature, depending on sign and notational conventions. In the present study, we use the conventions of Refs.~\cite{Dokshitzer:1995fv,Peigne:2023iwm,Peigne:2024srm,Cougoulic:2025wpl}.

The Kronecker deltas in quark and gluon indices are defined pictorially by edges: 
\bea
i \, \Idq{1} \, j  \ = \  \delta^{j}_{\ i} \ ; \ \ \ a\, \Idg{1} \, b \ = \ \delta_{ab} \ ,  
\eea
where, for $\sun$, quark and antiquark color indices are denoted by $i,j \in \{ 1,...\,, \Nc\}$ and gluon indices by $a,b \in \{ 1, ...\,, \Nc^2-1 \}$\,. (The index of an outgoing quark or incoming antiquark is written as a raised index by convention.) 

The $\sun$ generators in the fundamental (quark), complex conjugate (antiquark), and adjoint (gluon) representations, are denoted respectively by $T_{\rm F}^a$, $T^a_{\overline{\rm F}}$, and $T_{\rm A}^a$. Their matrix elements are represented by the following elementary birdtracks: 
\be
\label{eq:basic-vertices}
\gqqVert (1) \!= (T_{\rm F}^a)^{j}_{\ i} \equiv (t^a)^{j}_{\ i} \ ; \ \ 
\gqbarqbarVert (1) \!= (T^a_{\overline{\rm F}})^{j}_{\ i} = (-t^a)^{i}_{\ j} \ ; \ \ 
\gggVert (1) \!= (T_{\rm A}^a)_{bc} = -if_{abc} \ , 
\ee
where $f_{abc}$ are the (totally antisymmetric) $\sun$ structure constants. 
In our convention the $\sun$ antiquark generator includes a minus sign compared to the quark generator, $T^a_{\overline{\rm F}} = -(T^a_{\rm F})^\intercal$~\cite{Dokshitzer:1995fv}.
As a result, {\it each} basic vertex in \eq{eq:basic-vertices} is antisymmetric with respect to the permutation of any two of its outer edges. For the normalization of quark generators, 
\begin{equation}
\label{eq:norm-conv}
\tr t^a t^b = T_F \, \delta^{ab} \ \ \Longleftrightarrow \ \ \trtt{0.15} \ = \ T_F \ \Idg{0.8} \ \ , 
\end{equation} 
we use the standard convention $T_F = \frac{1}{2}\,$. 

Quite obviously, by removing color indices from matrix elements, we get a pictorial representation of {\it color maps}. For example, the quark line $\,\Idq{1}\,$ denotes the identity map $\unit_{\rm F}: q \to q\,$, where $q$ denotes the quark vector space (carrying the fundamental representation of $\sun$). Similarly, the gluon line $\,\Idg{1}\,$ is the identity map $\unit_{\rm A}: g \to g\,$, where $g$ denotes the gluon vector space.  

In our convention, the so-called {\it invariance condition}~\cite{Cvitanovic:2008zz} reads
\begin{equation}
\label{eq:invariance-cond}
\ColorConservationA{0.2} \ \ + \ \ \ColorConservationB{0.2} \ \ + \ \ \ColorConservationC{0.2} \ \ + \ \  \cdots \ \ + \ \ \ColorConservationD{0.2} \ = \ 0 \ \ , 
\end{equation} 
where the circular blob stands for any $\sun$ invariant tensor,
and the coupling of the vertical gluon from below always comes with a plus sign, independently of the type of parton it couples to.\footnote{In the convention of Ref.~\cite{Cvitanovic:2008zz}, the minus sign of the antiquark generator is not included in the pictorial definition of vertices, and the attachment of the vertical gluon to an antiquark thus comes with a minus sign in eq.~\eq{eq:invariance-cond}.} 

Rewriting eq.~\eq{eq:invariance-cond} by drawing some of the outgoing partons as incoming partons, and noting that the coupling of the vertical gluon to any parton line, from above, changes sign when the vertical gluon is ``stretched'' such that it couples to that line from below, eq.~\eq{eq:invariance-cond} can be expressed in an equivalent form, referred to as {\it color conservation}. For example, for any color map $A: q \bar{q} g \to gggg\,$, we have
\begin{equation}
\label{eq:color-cons}
\colconsLeft{0.7}  \ \  = \ \ \colconsRight{0.7}  \ \ , 
\end{equation}
where a grey ellipse crossed by a set of parton lines denotes the sum of all couplings of the vertical gluon to those lines.

Finally, let's mention the Fierz identity which, under the normalization convention $T_F = \frac{1}{2}$ in  \eq{eq:norm-conv}, is generally written as (see \eg\ 
Ref.~\cite{Dokshitzer:1995fv} for a derivation of this identity) 
\begin{equation}
\label{eq:Fierz-standard}
\Fierzmapa{1} \ = \ \frac{1}{\Nc} \ \Fierzmapb{1} \ + \ 2 \ \Fierzmapc{1} \ \ . 
\end{equation}
By rotating each birdtrack of \eq{eq:Fierz-standard} by 90 degrees (clockwise), then swapping the two lower (implicit) color indices, and keeping in mind the antisymmetry of basic vertices under permutation of two external edges, the Fierz identity can also be written as follows: 
\begin{equation}
\label{eq:Fierz}
-2\ \Vqq{1}{0} \ = \ \Xqq{1}{0} \ - \frac{1}{\Nc}\ \Idqq{1} \ \ . \hskip 5mm
\end{equation}
\vspace{2mm}

\bibliography{mybib}
\bibliographystyle{JHEP}

\end{document}